\documentclass[letterpaper,twocolumn,10pt]{article}
\usepackage[dvipsnames]{xcolor}
\usepackage{usenix2019_v3}

\usepackage{tikz}
\usepackage{amsmath}

\usepackage{fontawesome}

\newcommand{\tightpar}[1]{{\smallskip \noindent\bf #1}}
\usepackage{longtable}
\usepackage{multirow}
\usepackage{tablefootnote}
\usepackage{booktabs}
\usepackage{color}
\usepackage{listings}
\usepackage{enumitem}
\usepackage{balance}
\setlist[itemize]{noitemsep,topsep=2pt,leftmargin=*}

\definecolor{editorGray}{rgb}{0.95, 0.95, 0.95}
\definecolor{editorOcher}{rgb}{1, 0.5, 0} 
\definecolor{editorGreen}{rgb}{0, 0.5, 0} 
\definecolor{btq}{rgb}{0.03, 0.91, 0.87} 
\definecolor{dtq}{rgb}{0.0, 0.81, 0.82} 
\definecolor{cdb}{rgb}{0.37, 0.62, 0.63} 
\lstdefinelanguage{js}{
	keywords={typeof, new, true, false, try, catch, function, return, null, catch, switch, var, if, in, for, while, do, else, case, break,let, const, throw, with, await},
	keywordstyle=\color{Maroon}\bfseries,
	ndkeywords={class, export, boolean, throw, implements, import, this},
	ndkeywordstyle=\color{darkgray}\bfseries,
	identifierstyle=\color{black},
	sensitive=false,
	comment=[l]{//},
	morecomment=[s]{/*}{*/},
	commentstyle=\color{darkgray}\ttfamily,
	stringstyle=\color{OliveGreen}\ttfamily,
	escapeinside={/*\#}{\#*/},	
	morestring=[b]',
	morestring=[b]",
	morestring=[b]`
}
\usepackage{float}

\newcommand{\noGadgets}{{11}}
\newcommand{\noApplications}{{15}}

\newcommand{\noRCEsText}{eight}

\newboolean{extendedVer}
\setboolean{extendedVer}{false}

\ifthenelse{\boolean{extendedVer}}{%
}{%
}

\begin{document}

\date{}

\title{\Large \bf Silent Spring: Prototype Pollution Leads to Remote Code Execution in Node.js}

\author{
    {\rm Mikhail Shcherbakov}\\
  KTH Royal Institute of Technology
    \and
   {\rm Musard Balliu}\\
  KTH Royal Institute of Technology
   \and
   {\rm Cristian-Alexandru Staicu}\\
CISPA Helmholtz Center for Information Security
} 

\maketitle

\begin{abstract}
  Prototype pollution is a dangerous  vulnerability affecting prototype-based languages like JavaScript and
  the Node.js platform.
  It refers to the ability of an attacker to inject properties into an object's root prototype at runtime and
  subsequently trigger the execution of legitimate code gadgets that access these properties on the object's prototype,
  leading to attacks such as Denial of Service (DoS), privilege escalation, and Remote Code Execution (RCE).
  While there is anecdotal evidence that prototype pollution leads to RCE, current research does not tackle the
  challenge of gadget detection, thus only showing feasibility of DoS attacks, mainly against Node.js libraries.
 
  In this paper, we set out to study the problem in a holistic way, from the
  detection of prototype pollution to detection of gadgets, with the ambitious goal of finding end-to-end exploits beyond DoS, in full-fledged Node.js applications.
  We build the first multi-staged framework that uses \emph{multi-label} static taint analysis to identify prototype pollution
  in Node.js libraries and applications,  as well as a hybrid approach to detect
  \emph{universal gadgets}, notably, by analyzing the Node.js source code.  We implement our framework on top of GitHub's static analysis framework CodeQL to find
  11 universal gadgets in core Node.js APIs, leading to code execution. 
  Furthermore, we use our methodology in a study of
  15 popular Node.js applications to identify prototype pollutions and gadgets. We manually exploit \noRCEsText{} RCE vulnerabilities in three high-profile applications such as 
  NPM CLI, Parse Server, and Rocket.Chat.  
  Our results provide alarming evidence that prototype pollution
  in combination with powerful universal gadgets lead to RCE in Node.js.

\end{abstract}

\section{Introduction}

In recent years we have seen a growing interest in running JavaScript outside of the browser. A prime example is Node.js, a popular server-side runtime that enables the creation of full-stack web applications.
Its package management system, NPM, is the world's largest software repository with millions of packages. Researchers have studied this ecosystem extensively to discover several security risks \cite{StaicuPL18,StaicuP18,StaicuSBPS19,duantowards,Li21,Xiao21,BrownNWEJS17,staicu2021bilingual},
showing that these risks are further exacerbated by the interconnected nature of the ecosystem~\cite{ZimmermannSTP19}. While most prior work focuses on libraries,
the problem of automatically detecting vulnerabilities in Node.js applications is still open.

Prototype pollution is a JavaScript-driven vulnerability that manifests itself powerfully in the Node.js ecosystem. The vulnerability is rooted in the permissive nature of the language, which allows the mutation of an
important built-in object in the global scope -- \texttt{Object.prototype} -- called the root prototype. JavaScript's prototype-based inheritance enables accessing this important object through the prototype chain.
Thus, attackers can instruct vulnerable code to mutate the root prototype by providing well-crafted property names to be accessed at runtime. As a consequence, every object that inherits from the root prototype, i.e.,
the vast majority of objects in the runtime, inherits the mutation on the root prototype, e.g, an attacker-controlled property. This vulnerability was first introduced by Arteau~\cite{arteau2018prototype},
showing that it is a widespread problem in Node.js libraries. Recently, Li et al.~\cite{Li21,Li22} explore static analysis to detect prototype pollution vulnerabilities using object property graphs.

The few prior works~\cite{kim2021dapp,Xiao21,Li21,Li22,Kang22} on prototype pollution consider a successful attack any mutation of the root prototype. An immediate consequence of such mutations is Denial of Service (DoS) due to the overwriting of important built-in APIs,
e.g., \texttt{toString}. By contrast, our work  studies the implications of prototype pollution beyond DoS. 
In particular, we propose a semi-automated approach for detecting 
Remote Code Execution (RCE)
vulnerabilities pertaining to prototype pollution. While there is anecdotal evidence about the possibility of such attacks\cite{rce-evidence,arteau2018prototype}, we are the first to propose a principled and systematic approach to detect them. Our key focus is on  gadget identification and end-to-end exploitation which
no prior work has addressed thoroughly.

Moreover, we note the important similarities between object injection vulnerabilities (OIVs)~\cite{Dahse14,ShcherbakovB21} and RCEs based on prototype pollution.
These attacks work in two stages: (1) there is an untrusted flow from an application's
untrusted entry points to an \emph{injection sink}, e.g., the property of an object; (2) there is a gadget that further propagates the attacker-controlled data 
from the injection sink to a
security-relevant \emph{attack sink}. 
In analogy, the attacker loads the gun in stage one (by placing the payload into the injection sink), while letting someone else (a gadget) pull the trigger in stage two
and carry out the attack (through an attack sink).
We propose calling the class of OIVs pertaining to prototype pollution, \emph{prototype-based object injection} vulnerabilities (POIV).

In statically-typed languages, OIVs are enabled by insecure deserialization, which allows instantiating objects of an unexpected type, 
thus triggering otherwise unused methods. Similarly, in a duck-typed language like JavaScript, if an attacker mutates the root prototype, 
they change the dynamic type of multiple objects in the runtime. This in turn activates otherwise unused code paths that correspond to 
the new type, e.g., object \texttt{foo} having a property \texttt{bar} defined. Thus, code reuse is done at a finer granularity and 
in a less localized manner in dynamically typed languages. We also note that attackers can mutate several properties at once, 
hence chaining gadgets in the fashion of property-oriented programming~\cite{Dahse14}.

Our technical contribution is a multi-staged framework that uses multi-label static analysis for detecting prototype pollution, and a hybrid solution, 
i.e.,  combining dynamic and static analysis,
for detecting  gadgets.
We observe that code patterns that lead to prototype pollution, i.e., injection sinks, are rather rare in real-world code. Thus, different from prior work, we propose tuning the static analysis for improved recall,
rather than precision.
Additionally, to emphasize the feasibility of the attack, we detect \emph{universal gadgets} in Node.js' source code, 
which can be exploited in a wide-range of applications as they come packaged with the Node.js runtime.

Drawing on security advisories~\cite{snyk}, we aggregate a set of 100 vulnerable packages, which we use to 
design and validate our pollution detection analysis. In comparison with the state-of-the-art tool ODGen~\cite{Li22},
we empirically show that one can significantly increase recall and scalability, while only paying a modest decrease in precision.

We then design and evaluate our novel gadget detection analysis against four widely-used APIs for handling code or command execution in Node.js. 
We find a total of \noGadgets{} gadgets that can be triggered during typical execution of these APIs. While some gadgets enable code injection directly, 
others allow attackers to load  arbitrary files from the disk into the runtime, by confusing the module resolution mechanism. We also conduct a quantitative study 
on packages to estimate the prevalence of these gadgets in the Node.js ecosystem. We believe that we are the first to show
evidence that control flow can be hijacked in this way in Node.js, further emphasizing the dangers of shipping unused code with 
applications~\cite{KoishybayevK20}.

Finally, we analyze \noApplications{} popular Node.js applications, reporting on the effort to finding RCE  with our methodology.
We identify \noRCEsText{} exploitable RCE vulnerabilities in highly-popular applications such as NPM CLI, Parse Server and Rocket.Chat. We have responsibly disclosed these critical vulnerabilities to developers and they are now fixed, acknowledging our contributions with a high-severity advisory (e.g., CVE-2022-24760) and bug bounties. 

Contrary to established recommendations, this work  embraces false positives. We show that a motivated attacker can sieve through the manageable amount of 
false positives to find critical zero-day exploits against well-tested, mature applications. We believe that vulnerability detection tools tuned for offensive 
security can afford this luxury due to the high return on investment provided by a single true positive.

In summary, the paper offers the following contributions:

\begin{itemize}
  \item We are the first to study the impact of prototype pollution vulnerabilities in full-fledged Node.js applications, beyond denial-of-service attacks.
  \item We present a principled approach for detecting RCE vulnerabilities that are enabled by prototype pollution.
  \item We show that our pipeline is directly applicable to real-world code: we find \noGadgets{} universal gadgets in Node.js' source code and 
  \noRCEsText{} RCEs in popular applications.
  \item We provide initial evidence that unused code shipped with the application, e.g., third-party dependencies, can be leveraged as part of code reuse attacks in Node.js.
\end{itemize}

\section{Context and Technical Background}

This section provides background information and discusses the targeted threat model.
\ifthenelse{\boolean{extendedVer}}{%
We refer to Appendix~\ref{oiv-framing} to discuss POIVs in the general context of code-reuse vulnerabilities.
}{%
}

\subsection{Prototype-based OIV}

\emph{Prototypes} are a
key  feature to implement inheritance of JavaScript  properties and methods to  form a \emph{prototype chain}.
When creating an empty object, e.g., \texttt{const obj = \{\}}, it already contains many built-in properties and functions, for instance, the \texttt{toString} function.
When invoking \texttt{toString}  on an object, the runtime engine will first check  if the function is explicitly defined for the given object. If not, it will recursively look for its definition on the object's prototype chain.
Unfortunately, most objects share the same root prototype. For example, all objects created via the literal \texttt{\{\}} or the \texttt{new Object()} constructor share the same prototype unless it is explicitly overridden.
The following code snippet illustrates the problem:
\begin{lstlisting}
const o1 = {};
const o2 = new Object();
o1.__proto__.x = 42;
console.log(o2.x);
\end{lstlisting}
Although objects \texttt{o1} and \texttt{o2} are unrelated,  their prototype properties \texttt{\_\_proto\_\_}  point to the same object. In fact, if we add the
new property \texttt{x} to the prototype of object \texttt{o1} it will also affect object \texttt{o2}, resulting in a print of value \texttt{42} to the console.
Therefore, if we modify the root prototype shared by different objects, all these objects will reflect the modification.

We now explain the two stages needed to carry out a prototype-based attack that leads to code execution.

\tightpar{Stage 1: Polluting the prototype.}
Listing~\ref{lst:pp-template} shows a contrived example to illustrate key ingredients defining an \emph{injection sink} in a POIV.
We assume that the attacker controls all three arguments of function \verb|entryPoint|.
The first ingredient is an object that inherits a prototype that the attacker wants to pollute, as shown by the object in line 2,  
which inherits \verb|Object.prototype|.

\begin{lstlisting}[caption={Prototype pollution example},label={lst:pp-template}]
function entryPoint(arg1, arg2, arg3) {
  const obj = {};
  const p = obj[arg1];
  p[arg2] = arg3;
  return p; }
\end{lstlisting}

The second ingredient is the attacker-controlled access to the prototype property, as shown in line 3 via the bracket notation.
The attacker can pass \verb|__proto__| to \verb|arg1|  to store  \verb|Object.prototype| in  variable \verb|p|.
The last two ingredients require creating a target property in the prototype and assigning an attacker-controlled value. In fact, line 4 assigns an attacker-controlled value 
to a property of \verb|Object.prototype|.
Since the attacker controls \verb|arg2| and \verb|arg3|, they can write any value to any property. The JavaScript engine will create a new property, if such property does not exist.
In general cases, the attacker cannot fully control all  the ingredients, e.g.,  the property in \verb|arg2| or the value in \verb|arg3|.

An immediate effect of this vulnerable pattern is the attacker's ability to perform a DoS attack, e.g, by executing the function \verb|entryPoint("__proto__", "toString", 1);|
to alter the state to an unexpected integer value, i.e., \verb|Object.prototype.toString = 1|, thus, forcing an application that calls  \verb|toString()| to crash.

\tightpar{Stage 2: Executing the gadget.}
This stage requires identifying gadgets that contain insecure flows from injection sinks to \emph{attack sinks} that perform security-sensitive actions.

\begin{lstlisting}[caption={Gadget example},label={lst:gadget-template}]
const { execSync } = require("child_process");
function gadget(args, options) {
  const cmd = options.cmd || "cmd.exe /k";
  return execSync(`${cmd} ${args}`);
}
const args = ...;
gadget(args, {});
\end{lstlisting}

Consider the benign example in Listing~\ref{lst:gadget-template}, where a list of arguments \verb|args| and a command object \verb|options| is passed to a function \verb|gadget| with
the intention to execute command \verb|options| with arguments \verb|args|. The intended use of function \verb|gadget| is to either execute the command
that is specified via the property \verb|cmd| of the \verb|options| object
or execute the default command  \verb|cmd.exe|. However, since the
developer passed an empty object to function \verb|gadget| (line 7), the program is expected to execute the
default command, because \verb|options.cmd| is undefined (line 3).

Consider now an execution of the program in Listing~\ref{lst:pp-template} such that \verb|entryPoint("__proto__", "cmd", "calc.exe&");|
The attacker  manipulates the \verb|cmd| property of the root prototype,
causing  the undefined property \verb|options.cmd| to fall back to the value in the prototype chain. Hence, the attacker can control the command passed to \verb|execSync|,
which leads to code execution, launching a calculator via \verb|calc.exe&|.

\subsection{Threat Model}

Our threat model targets an attacker that controls the \emph{untrusted entry points} of a  Node.js application with the goal of
exploiting prototype-based OIVs to perform arbitrary code execution on the application. These untrusted entry points are 
application-specific, however, candidates include HTTP connections, untrusted database reads, and the like. We also consider a weaker threat model
targeting only \emph{universal} gadgets that occur in the source code of Node.js. Because these gadgets appear in code that executes with the Node.js runtime,
they are available for exploitation in any Node.js application. For this threat model, we assume that the attacker controls the injection sinks pertaining to
the execution of a gadget.

\section{Overview}\label{sec:overview}
This section provides an overview of our multi-staged analysis framework, illuminating on the key challenges in detecting and exploiting prototype-based object injection vulnerabilities. We use our newly-detected vulnerability in
NPM CLI to illustrate the complexity and challenges of such an endeavor.
NPM CLI\cite{npm-cli} is the command line client that allows developers to install and publish packages in NPM registries.
It comes bundled with the Node.js runtime and  consists of 713,648 lines of code.

\begin{figure}[t] 
  \includegraphics[width=8.5cm]{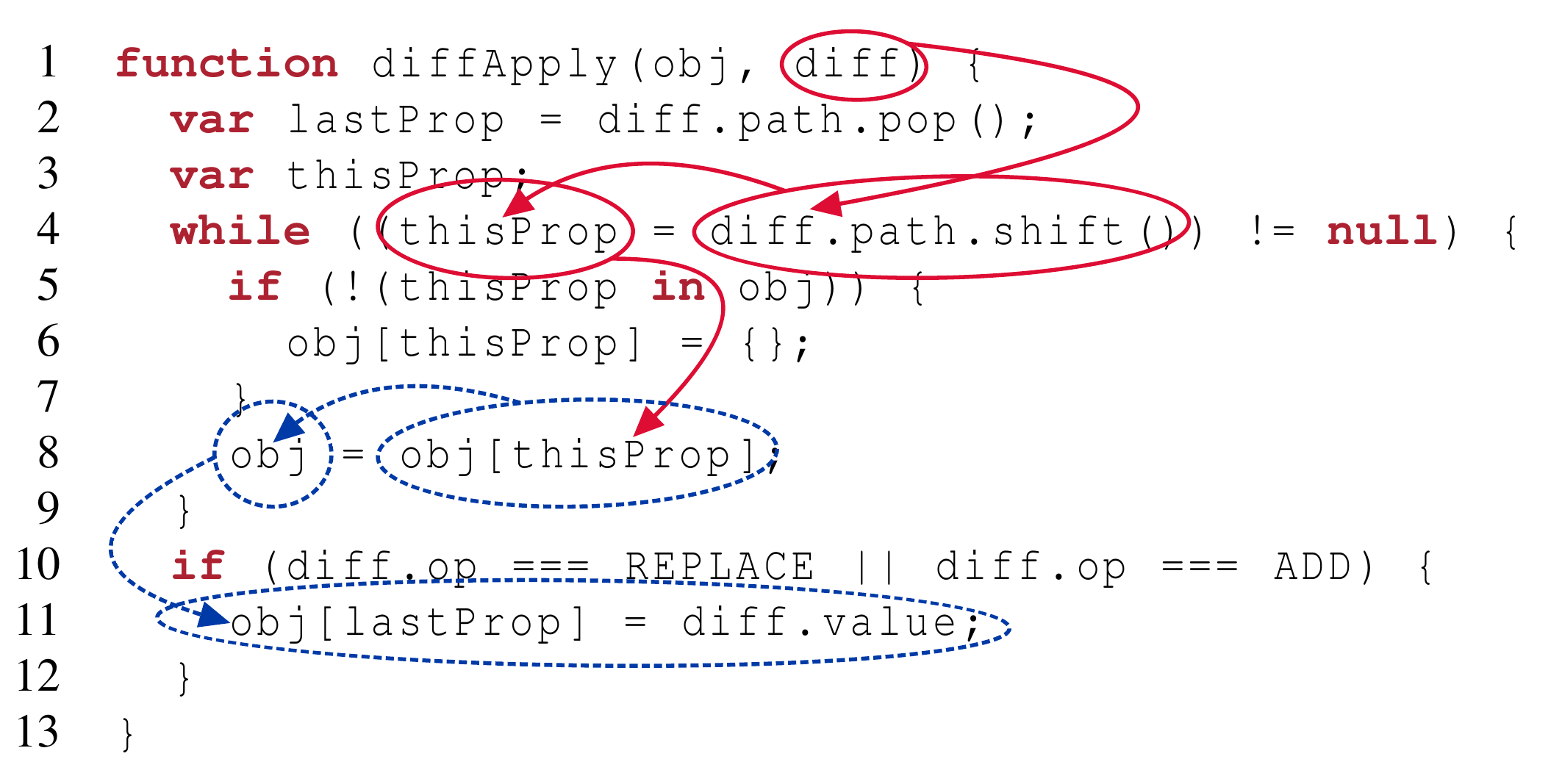}
  \vspace*{-.8cm}
  \caption{Injection sink in NPM CLI}
  \label{fig:pp-npm}
\end{figure}

\tightpar{Detecting prototype pollution.}
Figure~\ref{fig:pp-npm} shows the simplified code fragment of the function \texttt{diffApply} from NPM CLI's codebase, which is subject to prototype pollution.

The function takes the array \verb|path| from the attacker-controlled parameter \verb|diff| and calls the built-in function \verb|shift()|
that returns the first element of the array. The data flow then goes through the loop  storing a property value to the variable \verb|obj|  (red line).
Because the attacker indirectly controls the property name \verb|thisProp| in line 8, the property read  allows them to access the object's root prototype by
setting \verb|thisProp| to \verb|__proto__|.
Subsequently, the attacker can  assign any value to any property of the root prototype as illustrated by the assignment in line 11. As a result, the attacker has full control
of the injection sink denoted  by the blue dotted lines.  For instance, the function call \verb|diffApply({}, {path: ['__proto__', 'env'], value:| \verb|'payload', op: ADD})|
injects into \verb|Object.prototype|  the environment property \verb|env|  with payload \verb|payload|.

This code fragment illustrates the challenges that a static analysis should overcome. First, in contrast to standard taint analysis, injection sinks cannot be
identified syntactically as they require specialized data flow analysis that record accesses to object properties, as illustrated by the blue dotted line.
The analysis should identify attacker-controlled inputs that allow to control the prototype object, followed by uses of this prototype object
as a receiver in a property assignment~\cite{Li21}.
Second, the analysis should handle language constructs such as loops and model the JavaScript built-in functions, e.g., \verb|shift()| to correctly propagate data flows.
Third, given the size of the targeted codebases, the analysis should be scalable, seeking the sweet spot between precision and recall. While prior work achieves high precision, it reports low recall, thus increasing the possibility to miss flaws in real applications~\cite{Li22,Li21}. These requirements lead us to our
first research question: \emph{How to design and implement a scalable static analysis that  effectively identifies prototype pollution in real-world libraries and applications?} To answer this question we develop a multi-label static taint analysis, which we discuss in Section~\ref{sink-detection} and evaluate in Section~\ref{eval-detection}.

\begin{figure}[t] 
  \vspace*{-.5cm}
  \includegraphics[width=8.5cm]{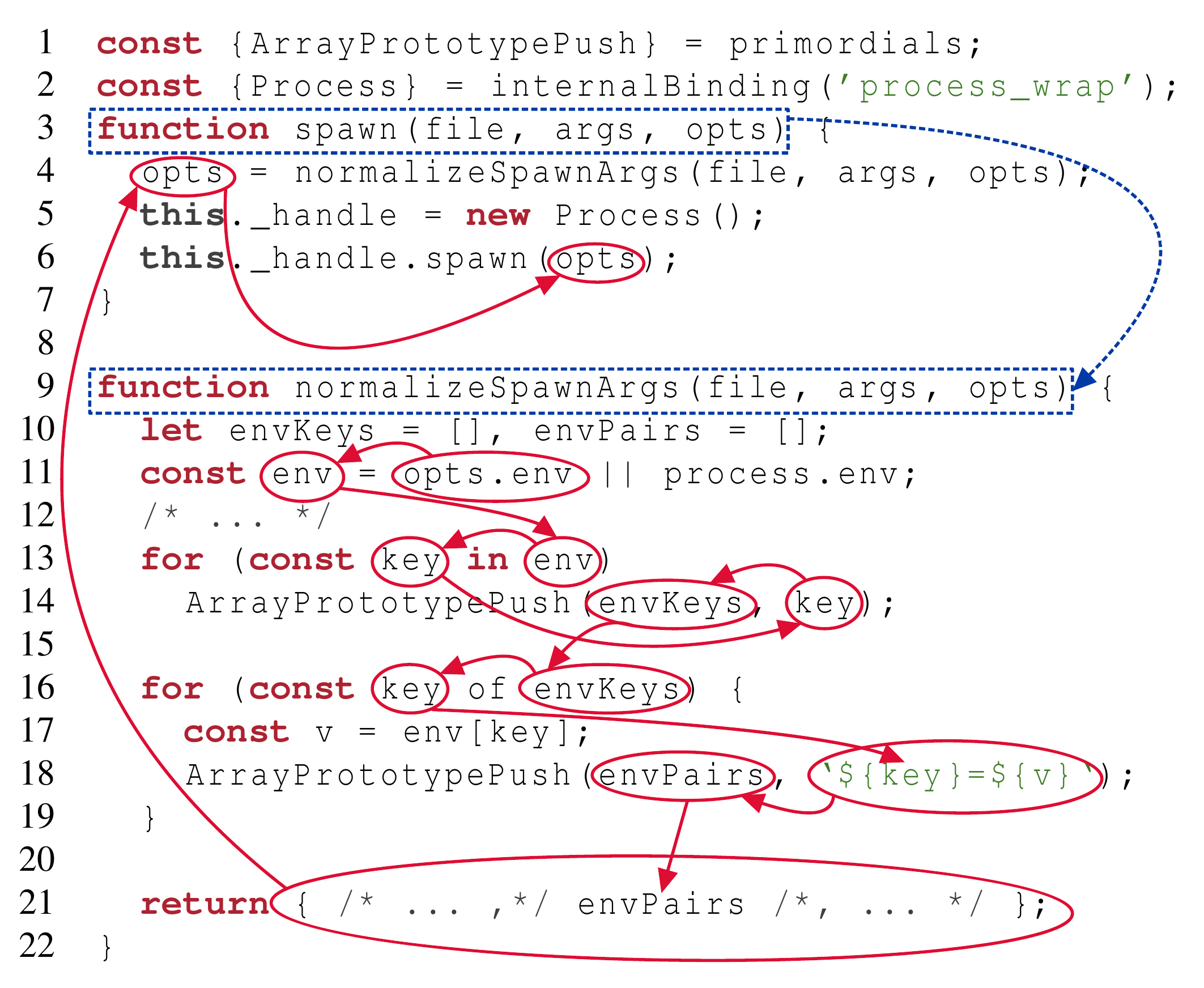}
  \vspace*{-.8cm}
  \caption{Universal gadget in Node.js standard library}
  \label{fig:gadget-nodejs}
\end{figure}

\tightpar{Detecting code gadgets.}
Recall that our threat model requires identifying code gadgets that read the attacker payloads from the injection sink and pass it into an attack sink.
Figure~\ref{fig:gadget-nodejs} shows a universal gadget we identified, stemming from the popular \verb|spawn| function of the Node.js standard library.
This function first calls \verb|normalizeSpawnArgs| and reads the value of property \verb|opts.env| in line 11. This optional parameter
contains key-value pairs of the environment variables of a new process. If a developer passes an object without property \verb|env|, the JavaScript
runtime will look up the property in the prototype chain.
Alternatively, attacker can inject the environment variable directly using the \verb|for..in| loop in line 13 to subsequently read it either from the \verb|opts.env|
or \verb|process.env| object in line 11.

The reader may at this point wonder about our second research question: \emph{How to identify universal properties reads such as} \verb|env|?
In fact, a prerequisite for gadget detection is to identify property reads that delegate the lookup of the property to the prototype chain, while
filtering out cases where the property is defined in the object itself. This is a complicated task for a static analysis, hence we use dynamic analysis instead.
\ifthenelse{\boolean{extendedVer}}{%
We discuss the details in Section~\ref{gadget-detection} and refer the reader to Appendix~\ref{appx-sources} for a, perhaps surprising, list of universal property reads on the root prototype.
}{%
We discuss the details in Section~\ref{gadget-detection}.
}

Further, the gadget contains intricate data flows from the property read in line 11 to the attack sink in line 6 as denoted by the red arrows.
Specifically, the \verb|for..in| loop enumerates the property names of the read object and passes them to an array through the \verb|ArrayPrototypePush| call.
This is an internal function that implements the semantics of \verb|Array.prototype.push| and subsequently
enumerates the \verb|envKeys| array, storing key-value pairs by the template literal (line 18)
and returning a new object with the property \verb|envPairs|. Therefore, an analysis should model the semantics of internal functions, template literals,
the \verb|for..in| and \verb|for..of| statements to propagate the attacked-controlled values properly.
Moreover, function \verb|spawn| (line 3) passes the modified object \verb|opts| to  method \verb|spawn| of the internal wrapper \verb|Process| (line 6) that is implemented
in the C++ component of Node.js. This method corresponds to the actual attack sink. Specifically, if an attacker uses \verb|{GIT_SSH_COMMAND: 'calc&'}| as \verb|payload| for
function \verb|diffApply|, they can simply wait for an invocation of the attack sink \verb|spawn| from the git command.
The latter uses the specified command from the environment variable \verb|GIT_SSH_COMMAND| when connecting to a remote system.
This leads us to our third research question: \emph{How to identify the attack sinks and data flows
  from universal property reads to these attack sinks in Node.js?} Gadget detection is a new challenge with no prior research, except for some evidence provided by the 
  practitioners' community~\cite{rce-evidence,arteau2018prototype}.  
  To address this question, we develop a taint-based static analysis that tracks flows from property reads to
attack sinks, which we discuss in Section~\ref{gadget-detection} and evaluate in Section~\ref{sec:universalgadgets}.

\tightpar{Putting it all together.}
The presence of prototype pollution and gadgets is not sufficient to carry out an end-to-end RCE attack. The attacker needs to identify application-specific untrusted entry points that enable the payload to reach the
injection sinks, and to subsequently propagate this payload to an attack sink via the gadget. This step requires us to combine data flow analysis with the call flow analysis,  starting
at untrusted entry points, while driving the payload to reach an attack sink. This leads to our final research question: \emph{How to identify public entry points and payloads to  demonstrate
  the feasibility of RCE attacks?} We use a combination of manual and automated analysis to drive the exploit towards success, as detailed in Section~\ref{end-to-end}
and evaluated in Section~\ref{exploit-apps}.

\section{Methodology}\label{sec:methodology}

\begin{figure}[t] 
  \includegraphics[width=8.5cm]{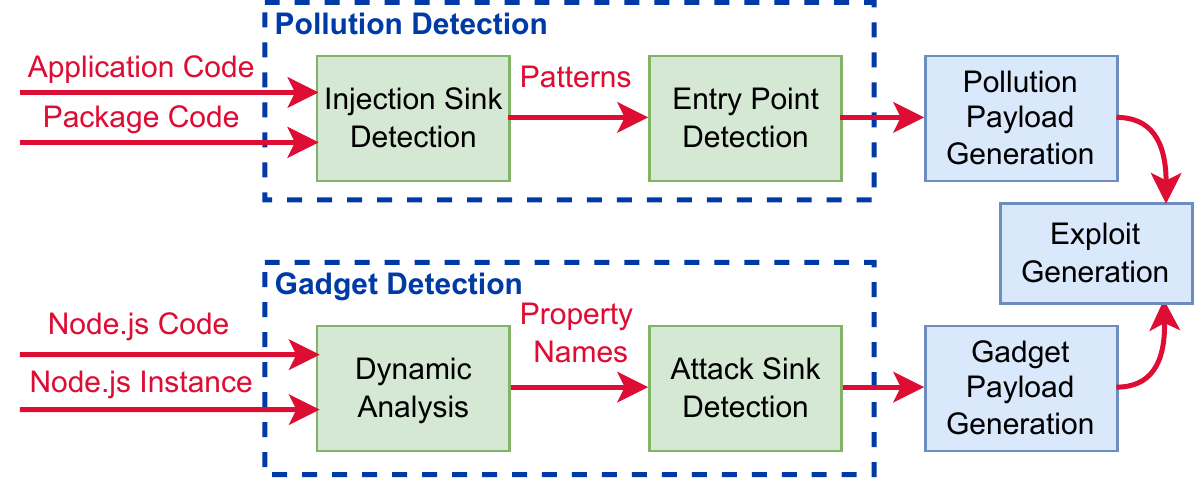}
  \vspace*{-.4cm}
  \caption{High-level workflow: automated steps (green) and manual steps (blue).}
  \label{fig:workflow}
\end{figure}

We present a semi-automated analysis framework for detecting and exploiting prototype-based vulnerabilities.
The framework is divided into three major steps: ($i$) automated prototype pollution detection; ($ii$) automated gadget detection; and ($iii$) manual  exploit generation for end-to-end attacks.
Figure~\ref{fig:workflow} illustrates the sequence of steps and their dependencies.

The prototype pollution detection step takes as input the code of an application or NPM package and performs a \emph{multi-label} taint-based static analysis. 
Subsequently, the analysis  reconstructs the call graph of the application to find entry points that reach the prototype pollution, thus facilitating the task of identifying attacked-controlled
entry points.
The gadget detection step implements a hybrid solution. A dynamic analysis first detects which properties can be actually polluted by executing Node.js APIs of interest in a testing environment that logs property accesses, ultimately
returning a list of accessed property names.
These property names, together with the source code of Node.js, are used as input to our second static analysis to identify (universal) gadgets in Node.js.
Each gadget includes an entry point that reaches a targeted property read and an attack sink that is called with values read from the target property.
The last step of the approach is the end-to-end exploit generation. This is a manual step that requires an investigation of the target application's workflow to validate
the exploitability of the detected prototype pollution and gadget to achieve code execution on the system.

\subsection{Prototype Pollution Detection}\label{sink-detection}
\tightpar{Multi-label taint analysis.} The detection of prototype pollution requires specialized data flow analysis that identifies injection sinks boiling down to the pattern \verb|obj[prototype][property] = value|.
We find these patterns by means of a flow- and context-sensitive \emph{multi-label} taint analysis. Specifically, we use two labels $input$ and $proto$ to capture the temporal relationship between (attacker-controlled)
property accesses in an object. We use label $input$  to mark parameters that are directly controlled by the attacker and label $proto$ to record that the attacker already controls the prototype of the labeled object.

The analysis works as follows: initially, it marks the  parameters of the analyzed function with the \emph{input} label. Then, it performs (standard) taint analysis propagating this label according to the JavaScript semantics
until it reaches a property read with a tainted value in the property name, e.g., \verb|obj[prototype]| with \verb|prototype| having label $input$.
This indicates that the attacker may control the property name and get access to the root prototype.
At this point, the label of the resulting property read, e.g., \verb|obj[prototype]|, is changed to the label $proto$ to record that the attacker can now control the prototype.
Subsequently, the analysis continues the taint propagation until it reaches a property assignment, e.g., \verb|obj[prototype][property] = value|,
where the object of the property assignment, i.e., \verb|obj[prototype]|, is marked with the \emph{proto} label, thus identifying the injection sink.  We note that this a general characterization of injection sinks, where the
attacker does not necessarily control the accessed property (\verb|property|) and the assigned value (\verb|value|), so long as they control the root prototype (\verb|prototoype|).
Because this setting is more difficult to exploit, our analysis supports a \emph{priority} mode to identify  attacker-controlled property names and values in a property assignment.
Specifically, it performs two additional operations to check that the property read (\verb|property|) and  the value (\verb|value|) are marked with  label \emph{input},
indicating that they  may be controlled by the attacker. As expected, these priority injection sinks  are an easier target for exploitation in practice.

Figure~\ref{fig:pp-npm} illustrates the multi-label taint analysis for the prototype pollution vulnerability in  NPM CLI. We consider the function \verb|diffApply| as target function and mark the parameters
with  label \emph{input}.
The red arrows depict the propagation of label \emph{input}. The parameter \verb|diff| is an object and the taint analysis passes the tainted label to all its properties. The method \verb|shift| is a built-in method that
returns the first element of the array. The static analysis models JavaScript standard built-in objects, and thereby, propagates the \emph{input} label to \verb|thisProp| in  line 4.
The next node of the data flow is the property read in  line 8, hence the analysis changes its label to \emph{proto}. The blue dotted lines then visualize the \emph{proto} label propagation.
The tainted value reaches the property assignment, and the algorithm reports this expression as the injection sink. This is also a priority sink because the parameters \verb|lastProp| and \verb|diff.value| in line 11
have label \emph{input}.

\textit{Methodology} 
We define the  (attacker-controlled) target functions in two ways: ($i$) a package's exported functions (dubbed Exported Functions) or ($ii$) any function of the analyzed codebase (dubbed Any Functions).
We use the first option for the package analysis only, assuming that the attacker controls any exported function and class of a package.
The second option allows us to analyze real-world applications with no knowledge of the application's entry points, which usually depend on the specific threat model.
We find this option  useful in practice to overcome inherent limitations of static analysis for JavaScript, which does not always support the correct label propagation, e.g., due to callbacks or dynamically-generated code.
In this case, the analysis allows us to detect injection sinks by propagating the \emph{input} label from the nearest function on the call graph. Yet, the semantic modeling of
built-ins is key to increasing the true positive rate.

Ideally, a taint analysis should provide precise and complete models of JavaScript constructs. CodeQL features many person-hour contributions into the modeling of built-in functions. Nonetheless, we observe that in practice these models are still insufficient. 
	Our approach relies on the ground truth provided by known vulnerabilities to improve the tool in modeling features that pertain to these vulnerabilities, thus reducing the number of false negatives.
	 Concretely, we review the CodeQL standard library to identify and fix language features,  e.g., Arrays and reflection calls (see Section~\ref{impl})  that affect the taint semantics for the considered packages. We applied this process iteratively to achieve high recall.

\tightpar{Entry point detection.} We propose a lightweight analysis to detect application-level entry points that may trigger the injection sinks.
This helps with applications that receive tainted data from external storage to  find the external action that
triggers the data acquisition from the storage. The static analysis first reconstructs a call graph where the functions with no callers are represented by nodes with outgoing edges only.
The algorithm considers such nodes as potential application entry points and reports the code paths  to the injection sink.

\textit{Summary}  This step provides information about the pollution patterns and  application's entry points for future manual validation and exploit generation.
We contribute five analysis variants: one analysis for entry point detection; two \emph{priority} analyses (for each type of target function) that report
injection sinks with all tainted ingredients; and two \emph{general} analyses (for each type of target function) that report injection sinks with a tainted receiver only.

\subsection{Gadget Detection}\label{gadget-detection}

\tightpar{Dynamic analysis.} 
We first parse the Node.js' source code and syntactically extract all directly-accessed properties.
The dynamic analysis  defines a custom handler with a property getter in \verb|Object.prototype| for each extracted property name.
We systematically analyze the Node.js API documentation to identify functions that potentially run processes or evaluate arbitrary code in the runtime.
We then invoke these  APIs to log their attempt of property reads from \verb|Object.prototype|, which result in 
reading uninitialized properties and getting the  value \verb|undefined|. This means that the values of these properties can be tampered via prototype pollution.
The dynamic analysis passes the collected property names to the next step.

\tightpar{Static analysis.} The analysis takes the Node.js' source code and the property names as input. 
The algorithm first performs the call flow analysis of  Node.js API functions, including information about aliases, ultimately 
allowing us to reconstruct a precise call graph of the analyzed functions.
We then use the call flow analysis to identify paths from any exported function to polluted property reads (identified by the dynamic analysis) and 
subsequently combine it with context-sensitive taint tracking to identify paths from these property reads to attack sinks, represented as tainted arguments to internal function calls. 
Specifically, the analysis propagates the taints on return values only for functions that are reached by the Node.js API on the analyzed call flow. 
Additionally, the analysis identifies affected exported functions that were not analyzed dynamically. For instance, 
the analysis of function \verb|spawn| reports a possible pollution of property \verb|env|. The static analysis shows the attack 
sinks that are affected by \verb|env| include additional Node.js API functions such as \verb|spawnSync|, \verb|exec| and \verb|fork|.

The taint analysis considers internal functions, i.e., functions  for 
which the analyzer cannot resolve the function body, as candidate attack sinks.  We conservatively cover all functions with no 
implementation in the codebase.
The taint analysis also uses  multi-labels. For property assignments, the algorithm propagates the taint label \emph{polluted} of the property 
and applies the new label \emph{receiver} to the receiver recursively.
For instance, if \verb|value| in the assignment \verb|obj.prop = value| has  label \emph{polluted}, then the analysis applies the \emph{receiver} label to \verb|obj| 
and the \emph{polluted} label to its property \verb|prop|. This is needed because we cannot enumerate all properties of an object when this object is used as parameter to an
attack sink.
Finally, the static analysis reports internal functions with no arguments and either \emph{polluted} or \emph{receiver} labels as attack sinks.

Figure~\ref{fig:gadget-nodejs} shows the analysis in action for  property \verb|env|. 
The blue dotted arrows illustrate the call flow analysis from the exported function \verb|spawn| to the first function call. The \verb|normalizeSpawnArgs| contains 
the  property read \verb|env| which is  the starting node of the taint analysis (red arrows).
Initially, the taint analysis propagates the label \emph{polluted} through the data flows. When the tainted value reaches the object creation statement 
in line 21, the analysis keeps the taint label for the property \verb|envPairs| and assigns the label \emph{receiver} to the created object.
This object is further propagated to the caller function and passed to the internal function \verb|_handler.spawn| in line 6, thus reporting  
\verb|_handler.spawn| as a candidate sink.

\subsection{Exploit Generation}\label{end-to-end}

Our approach relies on the human-in-the-loop model for exploit generation. For gadget exploits,
the information about attack sinks allows us to evaluate the impact of a polluted property and  filter out non-malicious sinks. 
The call flow and  taint analysis help to explore the code slice that reaches  the attack sink.
We use this information to generate a payload and test it on the detected Node.js APIs. We validate the detected  sinks and report 
new  gadgets for Node.js in Section~\ref{sec:universalgadgets}.

A security analyst first analyzes the  prototype pollution patterns to filter out false positives and non-executable cases in
the regular application workflow, e.g., patterns in testing code and development tools. For suspicious cases, the analyst uses the automatically-detected entry points to generate the first version of a payload
and validates it on the application. If an exploit fails, the analyst investigates the cause using other tools (e.g., a debugger) and modifies the payload. 

If the validation of the prototype pollution succeeds, then the next step is to search for gadget triggers.
We extend the universal gadget entry points (e.g., \verb|spawn|) with functions that evaluate JavaScript code represented as strings (\verb|eval()|, \verb|new Function()|, \verb|new vm.Script|)
and provide a call graph analysis for these calls. The analyst may use the call graph analysis to detect calls to these functions as well as the application's entry points that reach these calls.

If the analyst detects a gadget trigger, they need to validate that it is executed after the injection sink  and then generate a payload that pollutes the required properties.
If  code evaluation function is detected, the analyst investigates the preconditions for invoking it with attacker-controlled data. The input data can be read from the polluted property,
or the function's execution may be dependent on specific conditionals that use the polluted property. These steps lead to arbitrary code execution inside the Node.js instance.
We estimate the effort of using such exploitation model in a study in Section~\ref{exploit-apps}.

\section{Implementation}\label{impl}

CodeQL\cite{codeql} is a production-scale analysis engine  to perform semantics-based search on a target codebase, essentially by
treating code as data. The analysis first extracts a full hierarchical representation of  code (e.g., the AST) into a relational database. It then runs  analysis \emph{queries} against 
the database to compute result tuples, for instance, pairs of source locations and error messages for bug finding. 
CodeQL queries are written in a declarative, object-oriented logic programming language called QL,
which uses Datalog as underlying semantic model~\cite{avgustinov2016ql}.  
It also provides a  \emph{standard library} of queries that implement  control-flow and data-flow analyses, as well as support
for mainstream languages including JavaScript. The JavaScript model and the analyses are part of the  open-source QL standard library, making them amenable to extensions.

A key feature that we use in our analyses are \emph{path queries}  that describe the data flow between a source and a sink in the codebase. They support expandable taint tracking with 
the possibility of using multiple flow labels. 
This is essential to implement our analysis algorithms described in Section~\ref{sec:methodology}. 
Specifically, we develop the custom path queries for pollution and gadget detection.
We extend the taint tracking configuration to combine the call-flow and data-flow analyses, 
thus propagating tainted values through call flows in a context-sensitive way. 
This feature is essential for some of our analyses, e.g., to analyze entry points that receive tainted data from a database 
and not propagate the taint labels through code that is reachable from other entry points. 
We also model the array built-in functions \verb|reduce|, \verb|filter| and more, to correctly propagate tainted values via callback functions passed as arguments. 
This allows us to detect vulnerabilities that use \verb|reduce| in the injection sink.
We also resolve new functions created by \verb|bind| call to propagate taints from the provided values of the \verb|bind| arguments to the bound function parameters. 
Other changes include support for parameter passing via \verb|apply()| and \verb|call()| function calls, as well as the rest parameter syntax and the \verb|arguments| object. 
\ifthenelse{\boolean{extendedVer}}{%
We refer to Appendix~\ref{patt} for an example. 
}{%
}
We also improve the detection of exported functions of Node.js packages. Our analysis queries for pollution and gadget
detection follow the methodology described in Section~\ref{eval-detection} and are publicly available as complementary material~\cite{SilentSpringArtifacts}.

\section{Evaluation}

This section presents our experiments to validate the usefulness of our approach to detect and exploit POIVs.
We perform the experiments on an Intel Core i7-8850H CPU 2.60GHz, 16 GB of memory. The tool, the analysis results and data are available in the GitHub repository~\cite{SilentSpringArtifacts}.

\subsection{Evaluation of Prototype Pollution}\label{eval-detection}

This section evaluates the effectiveness of our tool to detect injection sinks, reporting on precision and recall.
While recent approaches already target this problem~\cite{kim2021dapp,Li21,Li22} for Node.js libraries,
our key contribution is scalability with low-to-moderate precision loss, while achieving high recall. In contrast to prior work on libraries, we
find that injection sinks are rare in real-world applications, motivating the need for high recall to identify exploitable vulnerabilities.

\tightpar{Benchmark.}
We compile an open-source dataset of 100 vulnerable Node.js packages, 
collected from the Snyk database~\cite{snyk}.
By studying the proof-of-concept exploit provided in the vulnerability report, we manually identify code locations (file name and line number) of injection sinks pertaining to the assignment of an attacker-controlled value to the polluted property.
We observe that some packages contain multiple exploitable injection sinks, which we also  add to our benchmark. 
This new dataset serves as ground truth to  evaluate the detection capabilities of static analyses. 
For comparison, we also consider the dataset of 19 packages provided by the state-of-the-art work ODGen~\cite{Li22}.

\tightpar{Setup.}
We use our benchmark to calculate the rate of true positives (TP), false positives (FP), and false negatives (FN) in an effort to identify the sweet spot between the precision and recall of the analysis.
The precision metric describes how well the tool identifies exploitable injection sinks, while 
recall  represents the fraction of real vulnerabilities reported by a tool. 
Following the methodology in Section~\ref{sink-detection}, we run our tool in four different modes with the goal of identifying the most effective approach for  detecting injection sinks in real-world applications.
Our benchmark shows that attackers can have different levels of control over the injection sinks. While in  general  it can be sufficient to control the injection of the root prototype only,
we notice that most exploits target injection sinks with attackers controlling both the name and value of a polluted property. Therefore, our tool distinguishes between the
two cases, respectively, denoted as \emph{General queries} and \emph{Priority queries}. Moreover, since our analysis considers transitive dependencies, we distinguish between target functions
considering \emph{Exported Functions} and \emph{Any Functions}, with the goal of identifying the best mode to analyze applications.

We also compare our results with three analysis queries which CodeQL recently made available publicly.
We consider these CodeQL queries as  baseline queries and run them on our benchmarks.
Moreover, we conduct a direct comparison with ODGen~\cite{Li22} on the dataset of 119 libraries. 

\tightpar{Results.}
We report the evaluation results in Table~\ref{tab:eval-server-side} in Appendix and here discuss only the precision and recall metrics 
in comparison with CodeQL's baseline queries and ODGen.

CodeQL provides three queries to detect prototype pollution, one of which yields no results, hence we discard it.
The remaining two queries detect vulnerabilities in 57 packages, with  47\% and  67\% precision and 42\% and 21\% recall, respectively.
While our analysis queries have been developed independently, our  main goal is to achieve high recall with good precision.
A fair comparison with the CodeQL baseline corresponds to our \emph{General} queries with \emph{Exported Functions}, which yields 35\% precision and 88\% recall.
The improved recall is due to better support for exported functions, array built-in functions, and complete semantic modeling of reflective invocations through
\verb|apply()|, \verb|call()| and \verb|build()| functions. These results confirm the challenge of statically analyzing data flows in  JavaScript without
precise models of the  language semantics and built-in functions.

Our second experiment is an evaluation of \emph{General} queries with \emph{Any Functions} as entry points. The analysis achieves 31\% precision and 97\% recall, producing 5 false negatives.
This false negatives are in packages such as \verb|Templ8| and \verb|total_js| with injection sinks into code that is generated dynamically via \verb|new Function()|, which CodeQL does not support.
The high recall shows that injection sinks appear in a few adjacent functions, which
reduces the risk of losing the taint marks because of missing models of built-in functions. However, precision deteriorates
because some detected patterns are not actually reachable from the library API with attacker-controlled arguments. We also notice the precision loss is much less than one would
expect from an analysis with the strong assumption that any function's arguments are attacker-controlled. We believe this is due to the shape of injection sinks requiring patterns that are not
very common in real-world code (see Section~\ref{sink-detection}). While 31\% precision in aggregate results is not ideal, our analysis produces less than 10 false positives for 90\% of the benchmarks.

Our third  experiment is the evaluation of \emph{Priority} queries with \emph{Any Functions} as entry points. In this setting, the attacker controls the name and value of the polluted property, thus it can leverage
any existing gadget. The analysis achieves 40\% precision and 93\% recall. The additional restrictions on  arguments increase the precision metric and keep high recall.
Because the analysis starts from any function and does not require specifying the entry points, we can easily apply it to real-word application analysis.
We identify this analysis query as the sweet spot between precision and recall, and use it to detect vulnerabilities in real applications (Section~\ref{exploit-apps}).

Our final experiment is a direct comparison with  ODGen~\cite{Li22}.   ODGen's analysis corresponds to our \emph{General} queries with \emph{Exported Functions}. ODGen is tailored towards high precision, while the authors recognize the need for high recall. In fact, our experiment shows that ODGen achieves 100\% precision and 50\% recall on the dataset of 19 libraries, 
while our analysis achieves  95\% precision and 95\% recall 
\ifthenelse{\boolean{extendedVer}}{%
(see Table~\ref{tab:eval-odgen}).
}{%
(see the evaluation results in~\cite{SilentSpringArtifacts}).
}
Nonetheless, ODGen detects vulnerabilities in 17 out of the 19 libraries, but fails to detect some variants of these vulnerabilities. We further evaluate ODGen on our dataset of 100 packages to find that it achieves 87\% precision and 33\% recall. 

\subsection{Gadget Detection}\label{gaddet}

We evaluate the feasibility of our universal gadget detection analysis and discuss the most important gadgets. 
We run our analysis on Node.js version 16.13.1 and  exploit each gadget both on Linux and on Windows operating systems.

\subsubsection{Dynamic Analysis}

We  download the source code of Node.js and parse it to extract all directly-accessed properties. We obtain a total of 18,741 property names for the 
analyzed 
codebase~\cite{NodejsRepo}.
For each name, we  install a getter on \texttt{Object.prototoype} to detect any potential access to that property by Node.js' internals.

Subsequently, we exercise the APIs under test with typical inputs from the Node.js documentation, e.g., execute the \texttt{ls} command with \texttt{spawn} \cite{NodejsDoc}, and log any potential accesses observed by the getter. 
In total, we analyze three APIs, i.e., \texttt{child\_process.spawnSync}, \texttt{require}, and \texttt{vm.runInNewContext}, and obtain 10, 11, and 16 candidate properties, 
respectively. The usage of these properties is further analyzed in the Node.js' codebase, using static analysis.

We note that the inputs used for driving the dynamic analysis are by no means exhaustive. We probably cover only a small part of the target APIs in our tests, potentially missing property accesses that only happen when the API is invoked with certain arguments. Nonetheless, for such cases, the resulting gadgets would be of limited use, as they would require the target application to pass those exact arguments to trigger the gadget. Instead of being comprehensive in our test case, we focus on the typical usages of the target APIs, which we believe yields easy-to-trigger gadgets.

Given the low number of properties detected in this step, one could directly fuzz these properties and build proof-of-concept exploits. 
However, we further trace their usage inside the Node.js codebase to understand if they are exploitable.

\subsubsection{Static Analysis}

As discussed in Section~\ref{gadget-detection}, our approach takes the JavaScript source code of Node.js and the property names from the dynamic analysis phase as input, 
and reports a call chain to reach a property read
and a data flow from the property read to an internal function invocation. We
only analyze the JavaScript code from the folder \textit{lib} of the repository~\cite{NodejsRepo}. 
The analyzed codebase contains 70,493 lines of code (LOC).

In total, we identify 778 exported functions that reach the property reads (sources), and 342 in which values read from these properties flow into internal functions (sinks). 
\ifthenelse{\boolean{extendedVer}}{%
In Appendix~\ref{ap:st-analysis} we present the detailed results, consisting of exact number of sources and sinks extracted for each universal property. 
}{%
}
We note that while inspecting all these code locations rigorously requires a significant amount of manual effort, we opt for pragmatic exploration: we first analyze the sink 
and decide if the invoked API, usually a native binding to the C/C++ code, is a relevant injection sink. If so, we continue with inspecting the sources 
to see which JavaScript APIs we can use to reach a particular code location.

Let us consider the case of \verb|shell|, a universal property identified by our dynamic analysis. The static analysis identifies 8 sources, 
meaning that the reads of \verb|shell| are reached from eight Node.js exported functions, mostly from the file \texttt{lib/child\_process.js}. 
By propagating taints from all detected property reads, we identify 11 function invocations in which the tainted value leaves the JavaScript world. 
One of them is located in the file \texttt{lib/internal/child\_process.js} and is a call to the native \texttt{spawnSync} in the  C++ bindings. 
By studying the bindings and the way they are invoked, we conclude that the \texttt{shell} universal property is a candidate for developing a gadget.

We thus proceed to further study the  operations performed on the value stored in the universal property inside the Node.js codebase. CodeQL provides great support in this step, allowing us to jump at the relevant code locations where this value is read and then manipulated. We already know from the dynamic analysis step that the Node.js core performs a read from this universal property when the function \texttt{spawnSync} is invoked, but by running a call graph reachability analysis we identify four other APIs that reach one of the sources.

We build a simple test case to first pollute the \texttt{shell} property with the value \texttt{touch} and then invoke one of the affected JavaScript API, 
i.e., \texttt{spawnSync}. By  observing the side-effect of this test case, i.e., the file creation  in the current directory, 
we conclude that if an attacker can pollute  \texttt{shell}, the API under test uses its value as command, instead of the argument passed by developers. 
We next discuss this gadget and others.

\begin{table*}[h]
  \centering
    \resizebox{2.0\columnwidth}{!}{%
  \begin{tabular}{|c|c|l|l|c|}
    \midrule
    \textbf{ID} &    
    \textbf{Universal properties}                                       & \textbf{Trigger}                                     & \textbf{Impact}                                                   & \textbf{OS}  \\
    \midrule
    $G_1$ & \texttt{shell}, \texttt{env}                               & Call command injection API                  & Execute  an arbitrary command                            & L+W \\
    $G_2$ & \texttt{shell}, \texttt{env}                               & Call command injection API                  & Execute  an arbitrary command                            & L   \\
    $G_3$ & \texttt{shell}, \texttt{input}                             & Call command injection API                  & Execute  an arbitrary command                            & W   \\
    $G_4$ & \texttt{main}                                              & Import a package without a declared "main"  & Import an arbitrary file from the disk{\color{red}$^*$}  & L+W \\
    $G_5$ & \texttt{main}                                              & Require a package without a declared "main" & Require an arbitrary file from the disk{\color{red}$^*$} & L+W \\
    $G_6$ & \texttt{exports}, \texttt{1}                               & Require a file using a relative path        & Require an arbitrary file from the disk{\color{red}$^*$} & L+W \\
    $G_7$ & \texttt{'=C:'}                                             & Resolve a file path                         & Resolve the path to a different file                     & W   \\
    $G_8$ & \texttt{contextExtensions}                                 & Require a file using a relative path        & Overwrite global variables of the file                   & L+W \\
    $G_9$ & \texttt{contextExtensions}                                 & Compile function in a new context           & Overwrite function's global variables                    & L+W \\
    \midrule
    $G_{10}$ & \texttt{shell}, \texttt{env}, \texttt{main}                & Require a package without a declared "main" & Execute an arbitrary command                             & L+W \\
    $G_{11}$ & \texttt{shell}, \texttt{env}, \texttt{exports}, \texttt{1} & Require a file using a relative path        & Execute an arbitrary command                             & L+W \\
    \midrule
  \end{tabular}
  }
  \caption{A summary of the identified Node.js universal gadgets. For each gadget, we show  the properties that the attacker must pollute beforehand, the action that triggers the gadget, and the produced effect. The last column shows the operating system on which the gadget works: Linux (L), Windows (W), or both (L+W). {\color{red}$^*$} denotes gadgets for which we have a Windows variant that achieves arbitrary command execution using the SMB protocol.}
  \label{tab:gadgets-pocs}
\end{table*}

\subsubsection{Universal Gadgets}\label{sec:universalgadgets}

We open source all the detected gadgets for Node.js in a GitHub repository~\cite{SSPPGadgets}.
Table~\ref{tab:gadgets-pocs} overviews the gadgets for the target Node.js version. 
Some of the gadgets are OS-specific, while most of them run on both considered OSs. 
We  emphasize the diverse set of universal properties involved, showing that gadgets are not isolated buggy cases, but they are common place. 
These gadgets correspond to a handful of target APIs inside the Node.js core, but that a motivated attacker can probably find many more 
inside the codebase of a target application. Finally, as we discuss below, some gadgets allow arbitrary code execution with a relatively strong precondition, 
while others allow hijacking the control flow with a weaker precondition. More importantly, an attacker can combine two such gadgets to get the best of both worlds.

We now discuss some of our most important gadgets and their assumptions to be fulfilled. Let us consider an application that invokes the \texttt{execSync} API with a string literal:
\begin{lstlisting}
const { execSync } = require('child_process');
console.log(execSync('echo "hi"').toString());
\end{lstlisting}
This benign looking code prints the string \texttt{hi} in the console. Staicu et al.~\cite{StaicuPL18} report that such API calls are prevalent in the NPM ecosystem, but they consider safe all call sites with constants as arguments, like the one above. That is because they assume an attacker cannot manipulate the command's value as it is set to a fixed value by developers. We find that this assumption does not hold in the presence of prototype pollutions. 
If attackers can pollute arbitrary properties in the runtime, they can hijack both the command to be executed and its environment variables. 
Consider the polluted properties:
\begin{lstlisting}
Object.prototype.shell = "node";
Object.prototype.env = {};
Object.prototype.env.NODE_OPTIONS =            "--inspect-brk=0.0.0.0:1337";
\end{lstlisting}

They trick the benign code above into spawning a new Node.js process with the debug port open, acting as a reverse shell. 
This is because the polluted property \texttt{shell} overwrites the command given by developers and \texttt{env.NODE\_OPTIONS} is set as environment variable 
of the current process and subsequently copied to all children processes. 

The presented gadget affects all the APIs for command execution in Node.js: \texttt{spawn}, \texttt{spawnSync}, \texttt{exec}, \texttt{execSync}, \texttt{execFileSync}. 
A precondition for this attack is that the target command execution call site should not explicitly set an options argument, e.g., for an \texttt{execSync} call, there should be no second argument passed.
The existence of this  gadget implies that {every Node.js application that is vulnerable to prototype pollution and uses a command execution API after a pollution is 
vulnerable to remote code execution}.

Now consider an application that does not directly use such APIs in user-facing code. An attacker can still leverage code that is present on the machine to trigger a command execution API. We found three gadgets that  exploit the \texttt{require} and \texttt{import} methods. Consider the following example:
\begin{lstlisting}
Object.prototype.main = "./../../pwned.js"
// trigger call
require('my-package')
\end{lstlisting}
A precondition for this gadget  is that  \texttt{my-package}  does not have a \texttt{main} property defined in its \texttt{package.json}. 
If the \texttt{main} property of the root prototype is polluted, at require time, the value of this property is used for retrieving the code to be executed, 
instead of the legitimate code of the module. The attacker can thus indicate an arbitrary file on the disk to be loaded in the engine. 
In particular, they can specify a file that contains calls to command execution APIs. 
For example, the popular \texttt{growl} package~\cite{growlNPM} contains a file called \texttt{test.js} that invokes 
the package with different test values. Considering that \texttt{growl} uses \texttt{spawn} internally, the attacker can successfully trigger such APIs call 
by setting the \texttt{main} property to point to the \texttt{growl}'s test file. Moreover, we identified a file shipped with the NPM command line tool that 
can be used for the same nefarious purpose: \texttt{npm/scripts/changelog.js}.

To the best of our knowledge, the gadget above is the first evidence ever reported that shows that hijacking control flow through code reuse attacks is possible in Node.js.
This motivates the need for debloating techniques like Mininode~\cite{KoishybayevK20}. 

In addition to the already alarming findings, an attacker can combine the two gadgets discussed above to obtain a powerful universal gadget:

\begin{lstlisting}
// pollutions for the first gadget
Object.prototype.main = "/path/to/npm/scripts/changelog.js";
// pollutions for the second gadget
Object.prototype.shell = "node";
Object.prototype.env = {};
Object.prototype.env.NODE_OPTIONS =            "--inspect-brk=0.0.0.0:1337";
// trigger call
require("bytes");
\end{lstlisting}

When the \texttt{bytes} package is loaded, the first gadget instructs the engine to load the \texttt{changelog.js} file. This file in turn invokes \texttt{execSync},  which triggers the second gadget, starting a Node.js process with a debugging session.

Finally, let us present another gadget that lets attackers load arbitrary files into the engine. By polluting the root prototype's properties \texttt{1} and \texttt{exports}, an attacker can execute an arbitrary file from the disk when a relative path is loaded:
\begin{lstlisting}
let rootProto = Object.prototype;
rootProto["exports"] = {".":"./changelog.js"};
rootProto["1"] = "/path/to/npm/scripts/";
// trigger call
require("./target.js");
\end{lstlisting}

While performing relative path resolution, the \texttt{require} method checks if the target path points to an ES6 module. During this process, the polluted  property \texttt{1} is inadvertently read when applying a destructuring operator 
\ifthenelse{\boolean{extendedVer}}{%
(see Appendix~\ref{appx-sources} for a discussion of complex pollution sources) 
}{%
}
in the file \texttt{/internal/modules/cjs/loader.js}:

\begin{lstlisting}
const { 1: name, 2: expansion = "" } = StringPrototypeMatch(...) || [];
\end{lstlisting}

Thus, the attacker-controlled value is assigned as the target module's name. Thereafter, the \texttt{require} method wrongly concludes that 
the relative path \texttt{./target.js}  resolves to the attacker-controlled location \texttt{/path/to/npm/scripts/} and that the path corresponds to an ES6 module. 
The \texttt{exports} property is used to confuse the \texttt{require} method further by providing the entry point for this non-existing module. 
Although at the attacker-controlled target location, there is no \texttt{package.json} file present,  the \texttt{require} method still concludes that this is a valid module path.
 We note that this gadget is not portable to legacy Node.js versions, e.g., version 14.15.0. Thus, an important precondition for exploitation is that the target system must use a recent Node.js version.

We emphasize once again how dangerous the identified gadgets are. Many fairly-large applications would probably meet the preconditions for an RCE, once a prototype pollution is in place: (i) require a file using a relative path or a package with no \texttt{main} entry, and/or (ii) have a dependency that uses a command execution API when loaded. 

To further study the impact of our gadgets, we estimate the prevalence of their triggers in an experiment with the 10,000 most dependent-upon NPM packages. We measure that 1,958 have no \texttt{main} entry in their package.json ($G_4$, $G_5$, $G_{10}$), 4,420 use relatives paths inside require statements ($G_6$, $G_8$, $G_{11}$), and 355 directly use the command injection API ($G_1$, $G_2$, $G_3$). This indicates that many of our gadgets could be deployed against clients of these packages, once a pollution is in place. However, this is an upper bound on the actual prevalence of the gadgets because: (i) the attacker may have a hard time invoking the trigger's code through the public interface of the package, e.g., the code using the command injection API, (ii) some gadgets may not work out of the box because of side-effects in the target package, i.e., polluting the property \texttt{1} may have many unintended side-effects that can prevent the gadget from working, (iii) an attacker may find it difficult to deploy a pollution before the gadget, e.g., for the require gadgets, very often, the pollution needs to happen in the application's initialization phase.  Nonetheless, considering the power of these gadgets and their widely-available triggers, prototype pollution should be considered a critical security vulnerability in the current Node.js landscape.

\subsection{End-to-End Exploitation}\label{exploit-apps}

\begin{table*}[t]
  \centering
  \resizebox{2.1\columnwidth}{!}{%
  \begin{tabular}{|r|c|c|cc|cc|cc|cc|cc|cc|}
    \hline
    \multirow{2}{*}{\begin{tabular}[c]{@{}r@{}}\textbf{Application's Repository}\end{tabular}}  & \multirow{2}{*}{\textbf{Stars}} & \multirow{2}{*}{\textbf{Lines of code}} & \multicolumn{2}{c|}{\textbf{Total}} & \multicolumn{2}{c|}{\textbf{Exploitable}} & \multicolumn{2}{c|}{\textbf{Suspicious}} & \multicolumn{2}{c|}{\textbf{Testing Code}} & \multicolumn{2}{c|}{\begin{tabular}[c]{@{}c@{}}\textbf{Client-Side Code}\end{tabular}} & \multicolumn{2}{c|}{\begin{tabular}[c]{@{}c@{}}\textbf{False Positives}\end{tabular}}  \\ \cline{4-15}
                                                                                                &                                 &                                         & \multicolumn{1}{c|}{Cases} & Time   & \multicolumn{1}{c|}{Cases} & Time         & \multicolumn{1}{c|}{Cases} & Time        & \multicolumn{1}{c|}{Cases} & Time          & \multicolumn{1}{c|}{Cases} & Time                                                      & \multicolumn{1}{c|}{Cases} & Time                                                      \\ \hline
    typicode/json-server           & 57,257 & 2,374   & \multicolumn{1}{c|}{0}  &       & \multicolumn{1}{c|}{-} &                                          & \multicolumn{1}{c|}{-} &        & \multicolumn{1}{c|}{-} &       & \multicolumn{1}{c|}{-} &      & \multicolumn{1}{c|}{-} &         \\ \hline
    expressjs/express              & 54,883 & 14,450  & \multicolumn{1}{c|}{0}  &       & \multicolumn{1}{c|}{-} &                                          & \multicolumn{1}{c|}{-} &        & \multicolumn{1}{c|}{-} &       & \multicolumn{1}{c|}{-} &      & \multicolumn{1}{c|}{-} &         \\ \hline
    meteor/meteor                  & 42,673 & 202,213 & \multicolumn{1}{c|}{26} & 255   & \multicolumn{1}{c|}{0} &                                          & \multicolumn{1}{c|}{5} & 210    & \multicolumn{1}{c|}{4} & 10    & \multicolumn{1}{c|}{8} & 5    & \multicolumn{1}{c|}{9} & 30      \\ \hline
    strapi/strapi                  & 40,724 & 168,998 & \multicolumn{1}{c|}{3}  & 5     & \multicolumn{1}{c|}{0} &                                          & \multicolumn{1}{c|}{0} &        & \multicolumn{1}{c|}{0} &       & \multicolumn{1}{c|}{0} &      & \multicolumn{1}{c|}{3} & 5       \\ \hline
    TryGhost/Ghost                 & 38,944 & 125,696 & \multicolumn{1}{c|}{4}  & 55    & \multicolumn{1}{c|}{0} &                                          & \multicolumn{1}{c|}{1} & 50     & \multicolumn{1}{c|}{0} &       & \multicolumn{1}{c|}{2} & 3    & \multicolumn{1}{c|}{1} & 2       \\ \hline
    hexojs/hexo                    & 33,666 & 21,073  & \multicolumn{1}{c|}{1}  & 40    & \multicolumn{1}{c|}{0} &                                          & \multicolumn{1}{c|}{1} & 40     & \multicolumn{1}{c|}{0} &       & \multicolumn{1}{c|}{0} &      & \multicolumn{1}{c|}{0} &         \\ \hline
    sahat/hackathon-starter        & 32,431 & 2,326   & \multicolumn{1}{c|}{0}  &       & \multicolumn{1}{c|}{-} &                                          & \multicolumn{1}{c|}{-} &        & \multicolumn{1}{c|}{-} &       & \multicolumn{1}{c|}{-} &      & \multicolumn{1}{c|}{-} &         \\ \hline
    koajs/koa                      & 31,910 & 4,596   & \multicolumn{1}{c|}{0}  &       & \multicolumn{1}{c|}{-} &                                          & \multicolumn{1}{c|}{-} &        & \multicolumn{1}{c|}{-} &       & \multicolumn{1}{c|}{-} &      & \multicolumn{1}{c|}{-} &         \\ \hline
    RocketChat/Rocket.Chat         & 31,059 & 242,949 & \multicolumn{1}{c|}{5}  & 1555  & \multicolumn{1}{c|}{1} & 1500 & \multicolumn{1}{c|}{3} & 50     & \multicolumn{1}{c|}{0} &       & \multicolumn{1}{c|}{1} & 5    & \multicolumn{1}{c|}{0} &         \\ \hline
    balderdashy/sails              & 22,085 & 24,445  & \multicolumn{1}{c|}{0}  &       & \multicolumn{1}{c|}{-} &                                          & \multicolumn{1}{c|}{-} &        & \multicolumn{1}{c|}{-} &       & \multicolumn{1}{c|}{-} &      & \multicolumn{1}{c|}{-} &         \\ \hline
    emberjs/ember.js               & 22,034 & 113,749 & \multicolumn{1}{c|}{6}  & 60    & \multicolumn{1}{c|}{0} &                                          & \multicolumn{1}{c|}{2} & 40     & \multicolumn{1}{c|}{1} & 10    & \multicolumn{1}{c|}{0} &      & \multicolumn{1}{c|}{3} & 10      \\ \hline
    fastify/fastify                & 21,043 & 37,049  & \multicolumn{1}{c|}{0}  &       & \multicolumn{1}{c|}{-} &                                          & \multicolumn{1}{c|}{-} &        & \multicolumn{1}{c|}{-} &       & \multicolumn{1}{c|}{-} &      & \multicolumn{1}{c|}{-} &         \\ \hline
    parse-community/parse-server   & 19,045 & 107,909 & \multicolumn{1}{c|}{7}  & 3225  & \multicolumn{1}{c|}{5} & 3220 & \multicolumn{1}{c|}{0} &     & \multicolumn{1}{c|}{0} &       & \multicolumn{1}{c|}{0} &      & \multicolumn{1}{c|}{2} & 5       \\ \hline
    docsifyjs/docsify              & 18,946 & 7,603   & \multicolumn{1}{c|}{0}  &       & \multicolumn{1}{c|}{-} &                                          & \multicolumn{1}{c|}{-} &        & \multicolumn{1}{c|}{-} &       & \multicolumn{1}{c|}{-} &      & \multicolumn{1}{c|}{-} &         \\ \hline
    npm/cli                        & 5,371  & 713,648 & \multicolumn{1}{c|}{15} & 603   & \multicolumn{1}{c|}{2} & 360                                      & \multicolumn{1}{c|}{6} & 230    & \multicolumn{1}{c|}{1} & 3     & \multicolumn{1}{c|}{0} &      & \multicolumn{1}{c|}{6} & 10      \\ \hline
  \end{tabular}
  }

  \caption{Evaluation results for the applications' analysis. \emph{Cases}  shows the number of detected cases of a certain category; \emph{Time} shows the time in minutes to manually classify and validate these cases.} \label{tab:eval-apps}
\end{table*}

We evaluate our approach on  popular Node.js applications from GitHub to validate its usefulness in a practical setting. 

\tightpar{Setup.}
We use the GitHub API to search for JavaScript repositories and order them by the number of stars. We then select for further analysis the top 14  web applications running on Node.js, as well as
NPM CLI, the JavaScript package manager, because it is installed on every machine with Node.js as default.
NPM CLI  is also the largest analyzed application in our dataset. We clone the GitHub repository of each application locally and perform the analysis against it.

\tightpar{Methodology.}
Following the workflow described in Section~\ref{sec:methodology}, we first run our \emph{Priority} query with  \emph{Any Functions} as entry points against a target application. The query reports the potential injection sinks and a list of the functions that pass tainted data to these sinks. The list contains functions that are actual entry points of the application and functions that take  data from the environment (e.g., a database) and pass it to the injection sink. For the latter, we perform a call flow analysis  to detect the application entry points.
Second, we manually classify all reported cases as either false positives or \emph{locally exploitable}. Based on the project structure, we also filter out  cases in testing  and client-side code. We discard these cases because the code does not execute on the server and cannot lead to RCE.
Third, we study the application's threat model to detect conditions for exploiting the remaining (locally exploitable) cases. This is a manual process that requires studying the documentation and code of the application. We match the entry points pertaining to  the threat model with the detected entry points leading to the injection sinks.
Fourth, we verify the matched entry points dynamically by deploying the application locally and generating a payload to pollute the \verb|toString| 
property. Whenever the payload fails, we rely on the  debugger by examining code transformations and validations along the path, and modifying the payload accordingly. Finally, once the pollution is confirmed, we search for the gadgets that may lead  to RCE, as described in Section \ref{gaddet}.
If the gadget can be triggered after the execution of the injection sink, we change the payload to pollute gadget-specific properties.

\tightpar{Results.}
Table~\ref{tab:eval-apps} presents the analysis results for 15 widely-used Node.js applications.  \emph{Total} provides the number of detected prototype pollutions 
in the application's codebase
and the total time for their manual analysis. The analysis finds cases in 8 applications, which we investigate and classify manually.  \emph{False Positives} contains the false positives
due to over-approximate  analysis; \emph{Client-Side} and \emph{Testing Code} show the cases  that do not execute on a server-side  directly.

We mark the remaining cases (column \emph{Suspicious}) for further investigation. 
Suspicious cases are locally exploitable patterns, i.e., they can be exploited if an attacker controls all function parameters.
We verified the suspicious cases to find \noRCEsText{} prototype pollutions (in NPM CLI, Parse Server and Rocket.Chat) that are exploitable according to the threat model of these applications.
We also found the gadgets that lead to RCE as explained below. As a sanity check, we  run the original CodeQL baseline queries for  NPM CLI and  
Parse Server applications, however, they do not detect exploitable prototype pollutions.

To estimate the manual effort, we track the time to verify the reported cases by one of the authors. A false positive  takes an average of 2.6 minutes  because the analysis affects  a small code fragment. Similarly,  non server-side code and testing code take on average 3.8 minutes  and 1.2 minutes, respectively. The analysis of suspicious  cases takes more time and depends on the quality of the documentation and application's code. The time in \emph{Suspicious} column includes the study of the threat model and the matching of detected entry points. The \emph{Exploitable} column includes the time to set up an application, debugging and verification of prototype pollution, search for gadgets, and combination of all attack ingredients. 
For example, most time for the Parse Server exploit was spend to find a race condition that triggers the injection and attack sinks in the correct order. 
For NPM CLI, a time-consuming task was to find a way to store the payload to NPM Registry via a malicious package and subsequently parse it during the package installation.
The analysis and exploitation of Rocket.Chat required an LDAP server setup that provides a payload to the injection sink, and the configuration of a custom synchronization with the LDAP server. This process is not fully described in the official documentation and required a lot of manual testing of various options.

We now describe the RCE exploits for two applications and refer to the extended material for full details \cite{SilentSpringArtifacts}.

\subsubsection{Parse Server RCEs}
Parse Server is an open source Backend-as-a-Service (BaaS) framework that provides REST APIs to object and file storage,
user authentication, push notifications, dashboard, and uses MongoDB or PostgreSQL as database. The Parse Server has pioneered  BaaS systems in 2011 and
has brought the serverless, low-touch deployment model to web and mobile backends.

\tightpar{Threat model.}
The Parse Server can be deployed as a self-hosted solution. In this scenario, an attacker can send any requests to the server, but cannot modify any settings on the server.
Therefore, we expect that an application must be secure in the default configuration.
In the second scenario, we consider the Parse Server as a part of cloud infrastructure, e.g., Back4App~\cite{back4app}.
The attacker can create their own account and become the administrator of that account. This allows the attacker to change some settings, for example, 
the webhook triggers.
This scenario puts any available configuration at risk for attacks including the default configuration.

\tightpar{Detecting sinks.}
Our static analysis framework detects 7 unique injection sinks. We marked 5 cases as suspicious by  manual validation. One of the suspicious cases is located in the sanitizer of database records as shown 
in Listing~\ref{lst:pp-parse-server}.

\begin{lstlisting}[caption={Injection sink in Parse Server},label={lst:pp-parse-server}]
function expandResultOnKeyPath(obj, key, res) {
  if (key.indexOf('.') < 0) {
    obj[key] = res[key];
    return obj;
  }
  const path = key.split('.');
  const firstKey = path[0];
  const nextPath = path.slice(1).join('.');
  obj[firstKey] = expandResultOnKeyPath(
    obj[firstKey] || {}, 
    nextPath, res[firstKey]);
  return obj;
}
\end{lstlisting}

This function can be abused to pollute \verb|Object.prototype|. If the attacker controls the input data and 
passes the value \verb|"obj.__proto__.evalFunctions"| to the parameter \verb|key| 
and the object \verb|{obj:{__proto__:{evalFunctions: 1}}}| to  \verb|result|,  then sanitization sets the new property \verb|evalFunctions| to Object's prototype.

Following our methodology, we perform a call flow analysis to detect entry points for the injection sink. A handler of the GET request triggers data reading from the database 
and then executes the vulnerable sanitizing code. Other detected injection sinks may be triggered via a PUT request 
by a payload delivered from a third-party webhook application.

In order to detect potential RCE gadgets, we search in Parse Server codebase for universal gadgets and functions that evaluate the code at runtime, e.g., \verb|eval|. 
The analysis reports a gadget using the  \verb|require| function, where an attacker can directly control its argument through 
a polluted property.   The analysis also reports an attack sink in the official MongoDB BSON parser~\cite{jsbson} that deserializes objects from a database, and 
can evaluate JavaScript code stored in this object. 
However, the code evaluation is possible only if we set the configuration parameter \verb|evalFunctions|, see Listing~\ref{lst:bson}. 
This option is not defined by default, but the attacker can  pollute the prototype and 
bypass the if-statement condition in line 5.

\begin{lstlisting}[caption={Attack sink in Parse Server},label={lst:bson}]
const evalFunctions = 
  options['evalFunctions'] == null 
  ? false 
  : options['evalFunctions'];
if (evalFunctions)
  eval(functionString);
\end{lstlisting}

\tightpar{Exploitation.}
The attacker should first pollute the prototype via the injection sink and then trigger the attack sink in a second request. 
A challenge to exploit prototype pollution  is  that the polluted property may break the application workflow. In this setting, the web request handler throws an 
exception whenever \verb|Object.prototype| is polluted. 
Thereby, the attacker cannot successfully handle the requests in the required order. However, we could bypass it using a \emph{race condition} in the application workflow.

Four of the RCE exploits for Parse Server use the same gadget and  attack sink in Listing~\ref{lst:bson} as follows: First, the attacker sends requests to store payloads in the database. Second, it sends the GET request to trigger the attack sink but 
delays its execution in the database until the next request. Third, the exploit sends the PUT request to trigger the injection sinks. 
Because the first request takes longer, a payload triggers the injection sink while another payload reaches the attack sink and executes arbitrary code. The fifth exploit adapts the
\verb|require| gadget discussed in Section~\ref{sec:universalgadgets}.

\subsubsection{NPM CLI RCEs}
NPM CLI\cite{npm-cli} is the command line client that allows developers to install and publish packages to NPM registries.
During a package installation,  NPM CLI puts modules in place so that Node.js can load them, manages dependency conflicts, and may run the pre- and post-install scripts from the package.

\tightpar{Threat model.}
The public NPM registry can be untrusted, e.g., by storing malicious packages.
Since it is a shell tool that is run on a developer's machine, RCE attacks have the highest impact. NPM CLI  has the option \texttt{-{}-ignore-scripts} to disable  running  scripts specified in \texttt{package.json} files.
Therefore, the threat model considers the arbitrary script execution that breaks out of the \texttt{-{}-ignore-scripts} flag as unintended RCEs.
We have the following constraint: the injection and attack sinks should be available during the  execution of the command that installs a malicious package.

\tightpar{Detecting sinks.}
The static analysis reports 15 unique injection sinks. We marked 8 cases as suspicious. Due to the restricted threat model, we then focus on matching the detected cases to the threat model.
When  NPM CLI installs the package, it parses the configuration
file \texttt{npm-shrinkwrap.json} from the package regardless of the option \texttt{-{}-ignore-scripts}. NPM CLI then invokes \verb|diff-apply| and \verb|copyPath| functions 
from the \verb|parse-conflict-json| package 
to parse the configuration file. 
Two of the suspicious cases are located in these functions.
Section~\ref{sec:overview} describes the injection sink in \verb|diff-apply| and the attack sink for the RCE exploitation. 
\ifthenelse{\boolean{extendedVer}}{%
Appendix~\ref{sec:npmrce22} shows the injection sink and the payload for the function \verb|copyPath|.
}{%
}
We verified manually that the exploitation in both cases leads to RCE.

\tightpar{Exploitation.}
The NPM CLI invokes the \verb|spawn| function to run the \verb|git| commands for git-located package dependencies. This happens after parsing the configuration files, and therefore, after the injection sink execution. 
The git supports the command execution via the environment variable \verb|GIT_SSH_COMMAND|. If this environment variable is set, git uses the specified command, instead of ssh,  to connect to a remote system. 
Thereby, the attacker can craft the package configuration file to initiate the call \verb|diffApply({}, {path:['__proto__','env'], value:| \verb|{GIT_SSH_COMMAND: 'calc &'}, op: ADD})| and wait for 
the \verb|spawn| invocation of the git command. This payload triggers arbitrary code execution, here launching a calculator.

\section{Related Work}

This section discusses closely related work targeting object injection vulnerabilities in general and
prototype pollution in particular. We also discuss related security analyses for the Node.js ecosystem and
client-side JavaScript security.

\tightpar{Prototype pollution vulnerabilities.}~The security community became aware of prototype pollution vulnerabilities in
2018 in a white paper of Arteau \cite{arteau2018prototype} which uses dynamic analysis to showcase feasibility in a number of
Node.js libraries as well as an end-to-end exploit in the Ghost CMS platform. The risks and the impact of prototype pollutions
has been mainly discussed in security practitioner forums~\cite{blackfan}, with the exception of a handful of recent research
papers~\cite{kim2021dapp,Xiao21,Li21,Li22,Kang22}. Notably, the work of Li et al.~\cite{Li21,Li22} proposes \emph{object dependence
  graphs} to statically find injection vulnerabilities in Node.js libraries, including prototype pollution. Object dependent graphs
allow identifying prototype injection sinks similar to our multi-taint analysis, though with higher precision due to the analysis of
branch conditions. By contrast, our approach trades precision for scalability to analyze fully-fledged applications and libraries.
 In addition, our key focus is on universal gadget identification and end-to-end exploitation which
no prior work has addressed systematically so far. Kim et al.~\cite{kim2021dapp} develop DAPP, a static analysis tool
to detect prototype injection sinks in Node.js libraries by means of pattern analysis. DAPP's lightweight analysis results in low precision
and recall, while focusing only on libraries. 
The recent work by Kang et al.~\cite{Kang22} explores prototype pollution on the client-side to exploit a range of vulnerabilities (XSS, cookie and URL manipulation) by using dynamic taint tracking.
Compared with static analysis,  dynamic analysis may miss some gadgets because of code coverage limitations, yet it 
can be helpful to validate the reachability of our injection and attack sinks, which we currently do manually.
Xiao et al.~\cite{Xiao21} study  hidden property attacks in Node.js applications, a type of vulnerability which is
related to prototype pollution.

\tightpar{Object injection vulnerabilities.}~ We classify POIVs in the general context of object injection vulnerabilities (OIVs).
Prior work studies OIVs targeting insecure deserialization by mean of static analysis in a variety of languages including Java~\cite{munoz2018serial,10.1145/2976749.2978361},
PHP~\cite{esser2010utilizing,Dahse14,Dahse14Usenix}, .NET~\cite{bh17,ShcherbakovB21},  and  Android~\cite{Peles15}.
The work of Dahse et al.~\cite{Dahse14,DahseH14} develops static analysis to  systematically detect OIV gadgets in PHP applications. Shcherbakov and Balliu~\cite{ShcherbakovB21} propose a static analysis
for detecting object injection patterns for .NET application, including the framework and libraries, and implement a tool called SerialDetector. Arguably, our work faces similar challenges with scaling the static analysis
to real-world languages, though in the more intricate context of JavaScript.

\tightpar{Node.js ecosystem security.}~There is an increasing interest in studying the security of Node.js, both in academia and in industry. Most prior work has concentrated on so-called software supply chain security, i.e., studying security problems that are prevalent in libraries: injections~\cite{StaicuPL18,Li22,GauthierHJ18}, hidden property abuse~\cite{Xiao21}, prototype pollution~\cite{Li21,Li22}, malicious packages~\cite{duantowards,ZimmermannSTP19}, running untrusted code~\cite{AhmadpanahHBOS21,VasilakisKRDDS18,VasilakisSNKKDP21}, ReDoS~\cite{StaicuP18,DavisCSL18,DavisSL21,LiuZM21}, code debloating~\cite{KoishybayevK20}. There is also initial evidence that these problems in libraries affect websites in production~\cite{Li21,StaicuP18}. We are the first to show the existence of universal gadgets in Node.js and to study the impact of prototype pollution, beyond denial-of-service attacks.

\tightpar{Static analysis for Node.js.}~Madsen et al.~\cite{MadsenTL15} propose augmenting call graphs with information about event propagation to find bugs in Node.js programs. Staicu et al.~\cite{StaicuPL18} advocate using intra-procedural data flow analysis to infer runtime policies for injection sinks. Nielsen et al.~\cite{NielsenHG19} introduce feedback-driven abstract interpretation for detecting injection vulnerabilities in Node.js code. More recently, Nielsen et al.~\cite{NielsenTM21} show how modular call graphs can be used to reduce false positives alerts in software composition analysis. Li et al.~\cite{Li21,Li22} propose using object dependency graphs for finding prototype pollution, injection, and path traversal vulnerabilities. We are the first to propose using static taint analysis for detecting universal gadgets.

\tightpar{Client-side JavaScript security.}~Lekies et al.~\cite{Lekies:2013:MFL:2508859.2516703} study XSS vulnerabilities on the web using fine-grained dynamic taint analysis. Hedin et al.~\cite{HedinBBS14} present JSFlow, a more sophisticated information flow analysis for detecting integrity and confidentiality problems in web applications. Recently, Lekies et al.~\cite{LekiesKGNJ17} discuss how script gadgets can be used to bypass existing cross-site scripting mitigation.
Roth et al.~\cite{Roth0S20} further study the effect of script gadgets on content security policies. Steffens and Stock~\cite{SteffensS20} present PMForce, a lightweight dynamic analysis augmented with forced execution for studying post message handlers. Khodayari and Pellegrino~\cite{KhodayariP21} propose JAW, a hybrid analysis tool based on code property graph, showing its usefulness by studying client-side CSRF vulnerabilities. None of the work above studies the relation between prototype pollution and injection vulnerabilities.

\section{Conclusion}

We presented the first principled study on the impact of prototype pollution vulnerabilities in Node.js. We propose a semi-automated approach for detecting end-to-end exploits, consisting of three phases: (i) static analysis for detecting pollutions, (ii) hybrid analysis for detecting gadgets, and (iii) static analysis with human-in-the-loop for developing end-to-end exploits. We apply our approach to large  codebases to find \noRCEsText{} exploitable RCE vulnerabilities directly enabled by prototype pollution, and eleven universal gadgets~\cite{SSPPGadgets} that are shipped with the Node.js runtime. Finally, we show that universal gadgets introduce a new threat in the Node.js ecosystem: hijacking the control flow of a program to (ab)use unused code available in the application's dependencies.

\paragraph{Acknowledgments}
Thanks are due to anonymous reviewers for the helpful feedback on this work. 
This work was partially supported by the Swedish Foundation for Strategic
Research (SSF) under projects CHAINS and Trustfull, Digital Futures, Google, and 
Wallenberg AI, Autonomous Systems and Software Program (WASP)
funded by the Knut and Alice Wallenberg Foundation.

\balance
\bibliographystyle{plain}
\bibliography{references}

\begin{thebibliography}{10}

\bibitem{back4app}
{Back4App}.
\newblock \url{https://www.back4app.com}.

\bibitem{jsbson}
{BSON Parser for node and browser}.
\newblock \url{https://github.com/mongodb/js-bson}.

\bibitem{blackfan}
{Client-Side Prototype Pollution and useful Script Gadgets}.
\newblock \url{https://github.com/BlackFan/client-side-prototype-pollution}.

\bibitem{codeql}
{CodeQL}.
\newblock \url{https://codeql.github.com}.

\bibitem{rce-evidence}
{Exploiting prototype pollution – RCE in Kibana (CVE-2019-7609)}.
\newblock
  \url{https://research.securitum.com/prototype-pollution-rce-kibana-cve-2019-7609}.

\bibitem{growlNPM}
{Growl - NPM. Growl support for Node.js}.
\newblock \url{https://www.npmjs.com/package/growl}.

\bibitem{NodejsDoc}
{Node.js documentation}.
\newblock \url{https://nodejs.org/api/child_process.html}.

\bibitem{NodejsRepo}
{Node.js JavaScript runtime v16.13.1}.
\newblock \url{https://github.com/nodejs/node/tree/v16.13.1/lib}.

\bibitem{npm-cli}
{NPM - a JavaScript package manager}.
\newblock \url{https://github.com/npm/cli}.

\bibitem{snyk}
Snyk.
\newblock \url{https://snyk.io}.

\bibitem{AhmadpanahHBOS21}
Mohammad~M. Ahmadpanah, Daniel Hedin, Musard Balliu, Lars~Eric Olsson, and
  Andrei Sabelfeld.
\newblock {SandTrap}: Securing {JavaScript}-driven trigger-action platforms.
\newblock In {\em {USENIX} Security Symposium}, 2021.

\bibitem{arteau2018prototype}
Olivier Arteau.
\newblock Prototype pollution attack in {NodeJS} application.
\newblock {\em NorthSec}, 2018.

\bibitem{avgustinov2016ql}
Pavel Avgustinov, Oege De~Moor, Michael~Peyton Jones, and Max Sch{\"a}fer.
\newblock Ql: Object-oriented queries on relational data.
\newblock In {\em 30th European Conference on Object-Oriented Programming
  (ECOOP 2016)}. Schloss Dagstuhl-Leibniz-Zentrum fuer Informatik, 2016.

\bibitem{BrownNWEJS17}
Fraser Brown, Shravan Narayan, Riad~S. Wahby, Dawson~R. Engler, Ranjit Jhala,
  and Deian Stefan.
\newblock Finding and preventing bugs in {JavaScript} bindings.
\newblock In {\em Symposium on Security and Privacy ({S\&P})}, 2017.

\bibitem{Dahse14Usenix}
Johannes Dahse and Thorsten Holz.
\newblock Static detection of second-order vulnerabilities in web applications.
\newblock In {\em {USENIX} Security 14}, pages 989--1003, 2014.

\bibitem{DahseH14}
Johannes Dahse and Thorsten Holz.
\newblock Static detection of second-order vulnerabilities in web applications.
\newblock In {\em {USENIX} Security Symposium}, 2014.

\bibitem{Dahse14}
Johannes Dahse, Nikolai Krein, and Thorsten Holz.
\newblock Code reuse attacks in {PHP:} automated {POP} chain generation.
\newblock In {\em Conference on Computer and Communications Security ({CCS})},
  pages 42--53, 2014.

\bibitem{DavisCSL18}
James~C. Davis, Christy~A. Coghlan, Francisco Servant, and Dongyoon Lee.
\newblock The impact of regular expression denial of service ({ReDoS}) in
  practice: an empirical study at the ecosystem scale.
\newblock In {\em Joint Meeting on Foundations of Software Engineering
  ({ESEC/FSE})}, 2018.

\bibitem{DavisSL21}
James~C. Davis, Francisco Servant, and Dongyoon Lee.
\newblock Using selective memoization to defeat regular expression denial of
  service ({ReDoS}).
\newblock In {\em Symposium on Security and Privacy ({S\&P})}, 2021.

\bibitem{duantowards}
Ruian Duan, Omar Alrawi, Ranjita~Pai Kasturi, Ryan Elder, Brendan
  Saltaformaggio, and Wenke Lee.
\newblock Towards measuring supply chain attacks on package managers for
  interpreted languages.
\newblock In {\em Network and Distributed System Security Symposium ({NDSS})},
  2021.

\bibitem{esser2010utilizing}
Stefan Esser.
\newblock {Utilizing Code Reuse/ROP in PHP Application Exploits}.
\newblock {\em Proceedings of the Black Hat USA}, 2010.

\bibitem{GauthierHJ18}
Fran{\c{c}}ois Gauthier, Behnaz Hassanshahi, and Alexander Jordan.
\newblock {AFFOGATO:} runtime detection of injection attacks for node.js.
\newblock In {\em {International Symposium on Software Testing and Analysis
  ({ISSTA})}}, 2018.

\bibitem{HedinBBS14}
Daniel Hedin, Arnar Birgisson, Luciano Bello, and Andrei Sabelfeld.
\newblock {JSFlow}: tracking information flow in {JavaScript} and its {APIs}.
\newblock In {\em Symposium on Applied Computing ({SAC})}, 2014.

\bibitem{10.1145/2976749.2978361}
Philipp Holzinger, Stefan Triller, Alexandre Bartel, and Eric Bodden.
\newblock An in-depth study of more than ten years of java exploitation.
\newblock In {\em Conference on Computer and Communications Security ({CCS})},
  pages 779--790, 2016.

\bibitem{Kang22}
Zifeng Kang, Song Li, and Yinzhi Cao.
\newblock Probe the proto: Measuring client-side prototype pollution
  vulnerabilities of one million real-world websites.
\newblock In {\em Network and Distributed System Security Symposium ({NDSS}
  2022)}, 2022.

\bibitem{KhodayariP21}
Soheil Khodayari and Giancarlo Pellegrino.
\newblock {JAW:} studying client-side {CSRF} with hybrid property graphs and
  declarative traversals.
\newblock In {\em {USENIX} Security Symposium}, 2021.

\bibitem{kim2021dapp}
Hee~Yeon Kim, Ji~Hoon Kim, Ho~Kyun Oh, Beom~Jin Lee, Si~Woo Mun, Jeong~Hoon
  Shin, and Kyounggon Kim.
\newblock Dapp: automatic detection and analysis of prototype pollution
  vulnerability in {Node.js} modules.
\newblock {\em International Journal of Information Security}, pages 1--23,
  2021.

\bibitem{KoishybayevK20}
Igibek Koishybayev and Alexandros Kapravelos.
\newblock Mininode: Reducing the attack surface of {Node.js} applications.
\newblock In {\em 23rd International Symposium on Research in Attacks,
  Intrusions and Defenses ({RAID})}, 2020.

\bibitem{LekiesKGNJ17}
Sebastian Lekies, Krzysztof Kotowicz, Samuel Gro{\ss}, Eduardo A.~Vela Nava,
  and Martin Johns.
\newblock Code-reuse attacks for the web: Breaking cross-site scripting
  mitigations via script gadgets.
\newblock In {\em Conference on Computer and Communications Security ({CCS})},
  pages 1709--1723, 2017.

\bibitem{Lekies:2013:MFL:2508859.2516703}
Sebastian Lekies, Ben Stock, and Martin Johns.
\newblock 25 million flows later: large-scale detection of {DOM}-based {XSS}.
\newblock In {\em Conference on Computer and Communications Security ({CCS})},
  pages 1193--1204, 2013.

\bibitem{Li21}
Song Li, Mingqing Kang, Jianwei Hou, and Yinzhi Cao.
\newblock Detecting {Node.js} prototype pollution vulnerabilities via object
  lookup analysis.
\newblock In {\em Proceedings of the 29th ACM Joint Meeting on European
  Software Engineering Conference and Symposium on the Foundations of Software
  Engineering}, ESEC/FSE 2021, page 268–279, New York, NY, USA, 2021.
  Association for Computing Machinery.

\bibitem{Li22}
Song Li, Mingqing Kang, Jianwei Hou, and Yinzhi Cao.
\newblock Mining {Node.js} vulnerabilities via object dependence graph and
  query.
\newblock In {\em {USENIX} Security Symposium}, 2022.

\bibitem{LiuZM21}
Yinxi Liu, Mingxue Zhang, and Wei Meng.
\newblock Revealer: Detecting and exploiting regular expression
  denial-of-service vulnerabilities.
\newblock In {\em Symposium on Security and Privacy ({S\&P})}, 2021.

\bibitem{MadsenTL15}
Magnus Madsen, Frank Tip, and Ondrej Lhot{\'{a}}k.
\newblock Static analysis of event-driven node.js javascript applications.
\newblock In {\em Proceedings of the 2015 {ACM} {SIGPLAN} International
  Conference on Object-Oriented Programming, Systems, Languages, and
  Applications, {OOPSLA} 2015, part of {SPLASH} 2015, Pittsburgh, PA, USA,
  October 25-30, 2015}, 2015.

\bibitem{bh17}
Alvaro Mu{\~n}oz and Oleksandr Mirosh.
\newblock {Friday the 13th json attacks}.
\newblock {\em Proceedings of the Black Hat USA}, 2017.

\bibitem{munoz2018serial}
Alvaro Mu{\~n}oz and Christian Schneider.
\newblock Serial killer: Silently pwning your java endpoints, 2018.

\bibitem{NielsenHG19}
Benjamin~Barslev Nielsen, Behnaz Hassanshahi, and Fran{\c{c}}ois Gauthier.
\newblock Nodest: feedback-driven static analysis of node.js applications.
\newblock In {\em Joint Meeting on European Software Engineering Conference and
  Symposium on the Foundations of Software Engineering, ({FSE})}, 2019.

\bibitem{NielsenTM21}
Benjamin~Barslev Nielsen, Martin~Toldam Torp, and Anders M{\o}ller.
\newblock Modular call graph construction for security scanning of node.js
  applications.
\newblock In {\em International Symposium on Software Testing and Analysis
  ({ISSTA})}, 2021.

\bibitem{Peles15}
Or~Peles and Roee Hay.
\newblock One class to rule them all: 0-day deserialization vulnerabilities in
  android.
\newblock In {\em {WOOT'15}}, 2015.

\bibitem{Roth0S20}
Sebastian Roth, Michael Backes, and Ben Stock.
\newblock Assessing the impact of script gadgets on {CSP} at scale.
\newblock In {\em Asia Conference on Computer and Communications Security,
  ({ASIA} {CCS})}, 2020.

\bibitem{ShcherbakovB21}
Mikhail Shcherbakov and Musard Balliu.
\newblock {SerialDetector: Principled and Practical Exploration of Object
  Injection Vulnerabilities for the Web}.
\newblock In {\em 28th Annual Network and Distributed System Security
  Symposium, {NDSS} 2021, virtually, February 21-25, 2021}, 2021.

\bibitem{SSPPGadgets}
Mikhail Shcherbakov, Musard Balliu, and Cristian-Alexandru Staicu.
\newblock {Server-Side Prototype Pollution Gadgets}.
\newblock \url{https://github.com/yuske/server-side-prototype-pollution}.

\bibitem{SilentSpringArtifacts}
Mikhail Shcherbakov, Musard Balliu, and Cristian-Alexandru Staicu.
\newblock {Silent Spring: Prototype Pollution Leads to Remote Code Execution in
  Node.js - Artifacts}.
\newblock \url{https://github.com/yuske/silent-spring}.

\bibitem{StaicuP18}
Cristian{-}Alexandru Staicu and Michael Pradel.
\newblock Freezing the web: {A} study of redos vulnerabilities in
  {JavaScript}-based web servers.
\newblock In {\em {USENIX} Security Symposium}, 2018.

\bibitem{StaicuPL18}
Cristian{-}Alexandru Staicu, Michael Pradel, and Benjamin Livshits.
\newblock {SYNODE:} understanding and automatically preventing injection
  attacks on {Node.js}.
\newblock In {\em Network and Distributed System Security Symposium ({NDSS})},
  2018.

\bibitem{staicu2021bilingual}
Cristian-Alexandru Staicu, Sazzadur Rahaman, {\'A}gnes Kiss, and Michael
  Backes.
\newblock Bilingual problems: Studying the security risks incurred by native
  extensions in scripting languages.
\newblock {\em arXiv preprint arXiv:2111.11169}, 2021.

\bibitem{StaicuSBPS19}
Cristian{-}Alexandru Staicu, Daniel Schoepe, Musard Balliu, Michael Pradel, and
  Andrei Sabelfeld.
\newblock An empirical study of information flows in real-world {JavaScript}.
\newblock In {\em 14th {ACM} {SIGSAC} Workshop on Programming Languages and
  Analysis for Security, {{PLAS}}}, 2019.

\bibitem{SteffensS20}
Marius Steffens and Ben Stock.
\newblock {PMForce}: Systematically analyzing postmessage handlers at scale.
\newblock In {\em Conference on Computer and Communications Security ({CCS})},
  2020.

\bibitem{VasilakisKRDDS18}
Nikos Vasilakis, Ben Karel, Nick Roessler, Nathan Dautenhahn, Andr{\'{e}}
  DeHon, and Jonathan~M. Smith.
\newblock Breakapp: Automated, flexible application compartmentalization.
\newblock In {\em Network and Distributed System Security Symposium, ({NDSS})},
  2018.

\bibitem{VasilakisSNKKDP21}
Nikos Vasilakis, Cristian{-}Alexandru Staicu, Grigoris Ntousakis, Konstantinos
  Kallas, Ben Karel, Andr{\'{e}} DeHon, and Michael Pradel.
\newblock Preventing dynamic library compromise on {Node.js} via {RWX}-based
  privilege reduction.
\newblock In {\em Conference on Computer and Communications Security ({CCS})},
  2021.

\bibitem{Xiao21}
Feng Xiao, Jianwei Huang, Yichang Xiong, Guangliang Yang, Hong Hu, Guofei Gu,
  and Wenke Lee.
\newblock Abusing hidden properties to attack the {Node.js} ecosystem.
\newblock In {\em {USENIX} Security Symposium}, 2021.

\bibitem{ZimmermannSTP19}
Markus Zimmermann, Cristian{-}Alexandru, Cam Tenny, and Michael Pradel.
\newblock Small world with high risks: {A} study of security threats in the npm
  ecosystem.
\newblock In {\em {USENIX} Security Symposium}, 2019.

\end{thebibliography}

\onecolumn
\section*{Appendix}

\ifthenelse{\boolean{extendedVer}}{%

\subsection{Object Injection Vulnerabilities}\label{oiv-framing}

Object Injection Vulnerabilities (OIVs) are an increasingly popular type of code-reuse vulnerability in the context of web applications.
They occur when an attacker can modify the properties of an object to abuse the data and control flow of the application.
OIVs enable attacker-controlled data to trigger the execution of legitimate code fragments (called gadgets) to perform malicious computations on the attacker's behalf.
For example, OIVs may arise during the deserialization of untrusted data from the client side, e.g., via  HTTP requests, when
reconstructing the object graph that is subsequently processed by the backend applications on the server side.
The following \emph{ingredients} are needed to exploit an OIV: ($i$) the attacker controls properties of an object to be instantiated, e.g., upon deserialization;
($ii$) the instantiated property affects execution of code gadgets in the application's scope;  ($iii$) there exists a big enough gadget space to find dangerous code fragments that
the attacker can chain to carry out, e.g., remote code execution. The attack works in two stages: (1) there is an untrusted flow from an application's
untrusted entry points to an \emph{injection sink}, e.g., the property of an object; (2) there is a gadget that further propagates the attacker-controlled data from the injection sink to a
security-relevant \emph{attack sink}. In analogy, the attacker loads the gun in stage one (by placing the payload into the injection sink), while letting someone else (a gadget) to pull the trigger in stage two
and carry out the attack (through an attack sink).  More generally, OIVs resemble second-order vulnerabilities, where an attacker first injects a value  through an injection sink and subsequently leverages
a read of that value to execute otherwise benign code paths (gadgets) that lead to the execution of an attack sink, possibly with attacker-controlled data from  the injection sink.
Existing work shows that OIVs are present in mainstream programming languages  like Java~\cite{munoz2018serial,10.1145/2976749.2978361}, JavaScript~\cite{LekiesKGNJ17},
PHP~\cite{esser2010utilizing,Dahse14,Dahse14Usenix}, .NET~\cite{foreshaw18,bh17,ShcherbakovB21},  and  Android~\cite{Peles15}.

\subsection{Non-trivial Gadget Sources}\label{appx-sources}

In this section, we compile a list of code patterns that imply a surprising read on the root prototype. Some of these patterns pose a great challenge for automatic static analysis of pollution gadgets. This list is not meant-to-be exhaustive, but instead to illustrate how difficult it is to write a comprehensive static analysis policy that can detect all property reads that lead to the root prototype.

We assume the attacker pollutes a property \texttt{x} of the root prototype and each of the pattern below reads this property. We remind the reader that after a successful pollution of the root prototype, every attempt to access a non-existent property \texttt{x} \emph{on every object} (including arrays) will lead to accessing the polluted property. Moreover, \texttt{x} is also available as a global variable to all programs, unless it is shadowed by other variables with the same name. Below, we discuss more subtle cases in which the property access is performed intrinsically by the language runtime.

To our surprise, important sources of user input, such as command line arguments and environmental variables can be influenced through a prototype pollution. If developers try to access a non-existent environmental variable \texttt{x}:

\begin{lstlisting}[language=js]
process.env.x
\end{lstlisting}

they would in fact read the attacker-controlled value \texttt{x}. Similarly, a direct read of a command line argument may lead to accessing attacker-controlled values:

\begin{lstlisting}[language=js]
process.argv[x]
\end{lstlisting}

Similarly, the module system in Node.js contains such careless property accesses. For example, every read on the built-in \texttt{exports} object can lead to reading polluted properties:

\begin{lstlisting}[language=js]
module.exports.x
\end{lstlisting}

Maybe more surprising, accessing property names on imported modules may also lead to polluted values:

\begin{lstlisting}[language=js]
const mod = require("fs");
mod.x;
\end{lstlisting}

This is especially problematic in the context of fast-evolving code, e.g., on NPM, where developers often check that a given method or property is available on an imported module.

Destructuring operators exhibit a surprising behavior, as well. Even when the operator is presented with a default value:

\begin{lstlisting}[language=js]
let {x = 12} = {};
\end{lstlisting}

the polluted value is assigned to \texttt{x}, instead of the default one.

For-in loops provide a convenient way o iterate through the keys of an object. Perhaps surprisingly for many readers, this code constructs considers all the inherited properties, hence, all the polluted property names as well:

\begin{lstlisting}[language=js]
for (a in arr) {
	// a will contain "x" in one iteration
}
\end{lstlisting}

Finally, \texttt{with} statements have the potential to introduce additional confusion:

\begin{lstlisting}[language=js]
let x = 12;
with({}) {
   // shadow the x above
   console.log(x);
}
\end{lstlisting}

Here, the polluted property is hiding a legitimate local variable, giving attackers enormous capabilities. For a long time now, this code construct was discouraged due to its complex semantics, thus, we believe such patterns are rather rare.

While we do not think that the above examples are bugs in Node.js, V8, or the ECMAScript standard, they are the enablers for powerful gadgets, like the universal ones described in this work. Thus, we advise language creators to avoid whenever possible such unprotected property reads, to reduce the prevalence of universal gadgets.

\subsection{NPM RCE II}\label{sec:npmrce22}

\tightpar{Injection sink.}
NPM CLI executes \verb|copyPath| functions 
from the \verb|parse-conflict-json| package 
to parse the configuration file. We demonstrate the source code of \verb|copyPath| to present the second vulnerability in NPM CLI:

\begin{lstlisting}[language=js]
const isObj = obj => obj && typeof obj === 'object'

const copyPath = (to, from, path, i) => {
  const p = path[i]
  if (isObj(to[p]) && isObj(from[p]) &&
      Array.isArray(to[p]) === Array.isArray(from[p]))
    return copyPath(to[p], from[p], path, i + 1)
  to[p] = from[p]
}
\end{lstlisting}

\tightpar{Exploitation.}
The NPM CLI invokes the \verb|spawn| function to run the \verb|git| commands for git-located package dependencies. This happens after parsing the configuration files, and therefore, after the injection sink execution. 
The git supports the command execution via the environment variable \verb|GIT_SSH_COMMAND|. If this environment variable is set, git uses the specified command to connect to a remote system. 
Thereby, the attacker can craft the configuration file as in the following example to trigger the injection sink and pollute the prototype. This payload triggers arbitrary code execution, here launching a calculator.

\begin{lstlisting}[language=js]
{ "obj": {
<<<<<<< 
    "__proto__": {"env": {"GIT_SSH_COMMAND": "calc &"} }
||||||| 
    "__proto__": {"env": {"GIT_SSH_COMMAND": ""} }
=======
>>>>>>>
}}  
\end{lstlisting}

\subsection{Advanced Prototype Pollution Pattern}\label{patt} 

In this section, we present an example of  prototype pollution in the  \emph{101} package from our benchmark to showcase the need for supporting JavaScript built-in functions and semantic models. The baseline CodeQL queries do not detect any vulnerability in this package, but our queries do because of extended support for JavaScript semantics in CodeQL standard library. Specifically,  the support for built-in functions  \verb|Array.prototype.reduce()| and \verb|Object.keys()| allows to detect such cases by the static analysis.

\begin{lstlisting}[language=js]
function reduceObject (target, source) {
  return Object.keys(source).reduce(function (obj, key) {
    if (isObject(obj[key]) && isObject(source[key])) {
      reduceObject(obj[key], source[key]);
      return obj;
    }
    obj[key] = obj[key] !== undefined ? obj[key] : source[key];
    return obj;
  }, target);
}
\end{lstlisting}

For a successful exploit,  the parameter \verb|target| should refer to an object with \verb|Object.prototype| and the value of \verb|source| should be controlled by an attacker. An example of successful exploit is the function call \verb|reduceObject({}, JSON.parse('{"__proto__":{"polluted":"yes"}}'))|.

To handle this code fragment properly, the static analysis should first propagate the tainted value from \verb|source| to the call \verb|Object.keys()|. 
The analysis should keep the taint mark for returned value of this function call following the modeled semantics. Moreover,
the static analysis then reflect the semantics of \verb|Array.prototype.reduce()| with good precision. The function takes the tainted array as a receiver and passes an element of the array to the parameter \verb|key| of the callback. The if-statement checks that both parameters have an object in the property \verb|key| and it recursively calls the function \verb|reduceObject|. 

In the next function call, \verb|source| still should be tainted because the parameter now refers to the property of the tainted object. In our example, the parameter \verb|target| refers to \verb|Object.prototype|  where the first key has the value \verb|__proto__| and should be marked by the corresponding label. The static analysis propagates the tainted value of \verb|source| to the if-statement again in the same way. It also propagates the tainted label from \verb|target| to \verb|obj| according to the semantics of \verb|Array.prototype.reduce()| (now from the second argument to the first parameter of the callback). 

Let us now consider another branch of the if-statement. The assignment expression in line 7 stores a value to a property of \verb|obj| where the name of the property and possible value \verb|source[key]| are tainted and therefore controlled by the attacker. Thus, the analyzer should report the assignment expression as an injection sink of the prototype pollution pattern.

\newpage
}{%
}

\subsection{Evaluation Results}

In Table~\ref{tab:eval-server-side}, we present the results of the evaluation of ODGen, the original CodeQL queries (\emph{Baseline queries}) and our custom queries (\emph{Priority queries} and \emph{General queries}) against our benchmark of 100 vulnerable NPM packages.

\footnotesize
\begin{longtable}[c]{|r|c|cccc|cccc|cccc|cc|}
  \hline
  \multirow{3}{*}{Package@Version}      & \multirow{3}{*}{LoC} & \multicolumn{4}{c|}{Baseline queries}                                                         & \multicolumn{4}{c|}{Priority queries}                                            & \multicolumn{4}{c|}{General queries}                                                                                                                                                                                                                                                                                                                                                                                                                                                                                                                                                                                                                                                                                                                                                                                                     & \multicolumn{2}{c|}{\multirow{2}{*}{ODGen}}  \\
\end{longtable}
\normalsize

\ifthenelse{\boolean{extendedVer}}{%
\newpage 
In Table~\ref{tab:eval-odgen}, we present the results of the evaluation of ODGen tool and our custom queries (\emph{Priority queries} and \emph{General queries}) against ODGen dataset of 19 vulnerable NPM packages.

\footnotesize
\begin{longtable}[c]{|r|c|cccc|cccc|cc|}
  \hline
  \multirow{3}{*}{Package@Version} & \multirow{3}{*}{LoC} & \multicolumn{4}{c|}{Priority queries}                                                       & \multicolumn{4}{c|}{General queries}                                                                                                                                                                                                                                                                                                                                                                                                                      & \multicolumn{2}{c|}{\multirow{2}{*}{ODGen}}                                                                  \\ \cline{3-10}
                                   &                      & \multicolumn{2}{c|}{\begin{tabular}[c]{@{}c@{}}Exported\\Functions\end{tabular}}            & \multicolumn{2}{c|}{\begin{tabular}[c]{@{}c@{}}Any\\Functions\end{tabular}} & \multicolumn{2}{c|}{\begin{tabular}[c]{@{}c@{}}Exported\\Functions\end{tabular}} & \multicolumn{2}{c|}{\begin{tabular}[c]{@{}c@{}}Any\\Functions\end{tabular}}                                                                                                                                                                                                              & \multicolumn{2}{c|}{}                                                                                        \\
\end{longtable}
\normalsize

In Table~\ref{ap:st-analysis}, we show the results of applying our CodeQL queries to the list of 37 universal properties inferred using dynamic analysis, in the previous phase of the gadget detection process.

\begin{table*}[h]
	\centering
  \resizebox{0.59\columnwidth}{!}{
	\begin{tabular}{|r|c|c|}
		\midrule   
		\textbf{Universal property} & \textbf{Number of sources} & \textbf{Number of sinks} \\
		\midrule
		cwd &	8 & 	2\\
		detached  &	0 & 	0\\
		uid  &	3  &	2\\
		gid  &	3  &	2\\
		shell & 	8  &	11\\
		argv0  &	9  &	16\\
		windowsHide  &	1  &	2\\
		windowsVerbatimArguments & 	0 & 	0\\
		env  &	136  &	52\\
		NODE\_V8\_COVERAGE  &	9  &	3\\
		timeout  &	21  &	6\\
		killSignal  &	8  &	2\\
		input  &	5  &	2\\
		output  &	4  &	2\\
		errmap 	 & 23  &	1\\
		main  &	5  &	6\\
		1 	 &127 & 	95\\
		2  &	28  &	14\\
		encoding  &	49 & 	20\\
		signal 	 &117  &	12\\
		href 	 &13 & 	20\\
		errno  &	21 	 &2\\
		error  &	8  &	8\\
		loaded 	 &0 	 &0\\
		NODE\_V8\_COVERAGE  &	9  &	3\\
		cachedData  &	15  &	4\\
		nullable  &	0  &	0\\
		name 	 &93 	 &29\\
		origin 	 &5  &	8\\
		codeGeneration  &	4  &	1\\
		microtaskMode  &	0 & 	0\\
		filename  &	18  &	9\\
		cachedData  &	15 & 	4\\
		contextName  &	3 & 	1\\
		contextOrigin  &	3  &	1\\
		contextCodeGeneration  &	0  &	0\\
    contextExtensions  &	7  &	2\\
    \midrule
    Total: & 778 & 342 \\
		\midrule                                                   
	\end{tabular}
  }
\caption{Complete results of our static analysis experiments for universal gadget detection. Sources and sinks represent unique code locations at which the universal property is read, or is passed into a native function, respectively.} \label{ap:st-analysis}
\end{table*}
}{%
}

\end{document}